\documentclass[12pt]{elsarticle}
\usepackage{fancyhdr}
\usepackage[hyperfootnotes=false]{hyperref}
\usepackage{graphicx}
\usepackage{comment}
\usepackage{dcolumn}
\usepackage{color}
\usepackage{longtable}
\usepackage{footmisc}
\usepackage{amsmath}
\usepackage{amsfonts}
\usepackage{enumerate}
\usepackage{braket}
\usepackage[algosection]{algorithm2e}
\usepackage{algorithmicx}
\usepackage{layouts}
\usepackage{tikz}
\usepackage{url}
\usepackage{caption}
\usetikzlibrary{positioning,decorations.pathreplacing}
\usepackage{mathabx}
\usepackage{enumitem}
\usepackage[margin=1in]{geometry}
\usepackage{amssymb}
\usepackage{verbatimbox}
\usepackage{footnote}
\makesavenoteenv{tabular}
\newcommand{\vect}[1]{\mathbf{#1}}
\newcommand{\reel}{\mathbb{R}}
\newcommand{\smalltt}[1]{\text{\tiny\texttt{#1}}}

\let\emptyset\varnothing
\journal{Computer Physics Communications}
\begin{document}
\begin{frontmatter}
\title{Dual vibration configuration interaction (DVCI).\newline 
{An efficient} factorization of molecular Hamiltonian for high performance infrared spectrum computation.}
\author[a]{Romain Garnier\corref{author}}
\cortext[author]{\textit{E-mail address:} rom1$\{dot\}$garnier$\{at\}$yahoo$\{dot\}$fr}
\address[a]{Queen's University of Belfast}
\begin{abstract}
Here is presented an original program based on molecular Schr\"odinger equations. It is dedicated to target specific states of infrared vibrational spectrum in a very precise way with a minimal usage of memory. 
An eigensolver combined with a new probing technique accumulates information along the iterations so that desired eigenpairs rapidly tend towards the variational limit.
Basis set is augmented from the maximal components of residual vectors that usually require the construction of a big matrix block that here is bypassed with a new factorisation of the Hamiltonian. The latest borrows the mathematical concept of duality and the second quantization formalism of quantum theory.
\end{abstract}
\end{frontmatter}
{\em Digital Object Identifier : \href{https://doi.org/10.1016/j.cpc.2018.07.008}{10.1016/j.cpc.2018.07.008}}~\newline~\newline
{\bf PROGRAM SUMMARY}~\newline~\newline
\begin{small}
\noindent
{\em Program Title:  Dual vibration configuration interaction (DVCI)}~\newline ~\newline
{\em Licensing provisions: \href{https://www.gnu.org/licenses/gpl-3.0.en.html}{GNU General Public License 3}.}~\newline~\newline 
{\em Programming languages: C/C++/Fortran}.~\newline ~\newline    
{\em Supplementary materials:\begin{enumerate} \item The sources of the code grouped in folder DualVCI.zip also available at \newline
 \url{https://github.com/4Rom1/DualVCI}\item The input files of examples treated in section \ref{Examples}. \end{enumerate}}~
{\em Nature of problem: High computational cost in vibration configuration interaction methods \cite{Christoffel1982,Thompson1980},
coming from the necessity to solve a large eigenvalue problem to acquire a good precision. The dimension of the matrix exponentially increases with the size of the studied molecule.}\\
  ~\\
{\em Solution method: The $A_k$ decomposition \cite{AK} completed by a meaningful error evaluation namely the residue $||HX-EX||$ minimised along iterations for specific targets given in input. The approximation space is generated in the same time as the residual vectors computed on the fly thanks to an adapted choice of excitations shaped on the Hamiltonian operator.}\\
\end{small}

\section{Introduction}
Nowadays, many devices are able to supply high quality spectrum measurements. However the interpretation of the samples remains a difficult task because the numerical accuracy that is possible to obtain is very limited for medium to large molecules ($> 6$ atoms).
In a typical resolution of vibrational Schr\"odinger equations in the Born-Oppenheimer frame, one can question the correctness of the model from two principal angles.
First, the quality of the results will be affected by the one of the Potential Energy Surface (PES).
This last point aside, there remains the validity of the numerical solutions when the PES is provided. That is the focus of the proposed method. 
 A prior analysis of the potential energy supplies the harmonic states that are not able to correctly describe complex combinations of translational and rotational motions of the nucleus. Although still accurately limited, the VSCF method \cite{Bowman1979,Bowman+1986} alone allows a better representation.
With the idea that these first estimations can be combined to give a more authentic description, the cheapest technique remains the perturbation theory \cite{Christiansen2003,Yagi2009,Respondek2009} that is known to struggle with strong resonances conventionally encountered in molecular spectroscopy \cite{Herman2013,Roth2009}. 

Vibration configuration interaction (VCI) method \cite{Christoffel1982,Thompson1980,Christiansen2007} permits a better precision with a much higher computational expense coming from the size of the variational space that is tenfold with the number of atoms of the molecule. {To avoid this bottleneck, contraction techniques have been proposed with MCTDH \cite{Multi-configurational1989,Worth2000}, followed by the Alternating Least Square (ALS) procedure \cite{Beylkin2005} formulated for wave function representation with the Vibrational Coupled Cluster (VCC) theory  \cite{OVE,Madsen2017}.}
{In the category of variational methods using perturbation criteria,} the $A_k$ decomposition \cite{AK} {originally} designed for electronic structure calculation and introduced in the VCI context with the Vibrational Multi Reference Configuration Interaction (VMRCI) \cite{Pfeiffer2014}, has been able to identify relevant sub-blocks of the Hamiltonian matrix thereby reducing the size of the system. {Analogous construction has been employed with PyVCI\_VPT2\cite{PyVCIVPT2,Sibaev2016}\footnote{{VPT2=Vibrational Perturbation Theory of order 2}} and} Adaptive-VCI (A-VCI) \cite{Garnier2016,Odunlami2017} to {respectively} access {the fundamentals and} smallest eigenvalues with a very good precision.
These last approaches showed promising results in term of size reduction, but still require to a posteriori determine  a big matrix block designed to improve the accuracy of the solutions.
In the present work, the non zeros of this sub-block are never collected and the number of operation to perform proper Matrix Vector Products (MVPs) is highly reduced thanks to a new factorisation of the Hamiltonian.
Beside, the expense of RAM is even more diminished because there is the possibility to constraint numerical accuracy on a few specified targets.
 In next sections are presented the context and the state of the art. The concept of duality intervenes all along the paper each time an association is made between a group of objects especially in section \ref{NewAlgo} explaining the theoretical aspects of the algorithm. Thereafter a description of the successive processed operations are followed by benchmarks. In this section, the method can provide comparable quality (i.e less than 1$ \ \rm cm^{-1}$ {deviation}) of a full A-VCI calculation, but with a memory consumption scaled down by more than a factor 15. Next, the list of input parameters of the program serves as a user manual. At the end, the conclusion also mentions a list of possible future developments.
\section{Context}
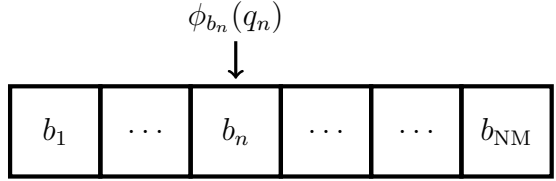
\begin{figure}[!htbp]
\center
\begin{tikzpicture}
 [box/.style={rectangle,draw=black, ultra thick, minimum size=1.2cm},]
\foreach \x/\y in {0/$b_1$, 1.2/$\ldots$, 2.4/$b_n$, 3.6/$\ldots$, 4.8/$\ldots$, 6/$b_{\mathrm{NM}}$}
       \node[box] at (\x,0){\y};
\draw[->,very thick] (2.4,1.2) --  node[above,yshift=2mm]{$\phi_{b_n}(q_n)$} (2.4,.7);
\end{tikzpicture}
\captionsetup{justification=centering}
\caption{Multi index array identified with a basis function. \newline Each index is an Hermite function degree corresponding to an harmonic quantum level.} 
\label{Fig:KInd}
\end{figure}

For a molecule composed of $N_A$ atoms, one considers NM dimensionless normal coordinates \cite{Wilson1955} $\vect{q}=(q_1,q_2,q_3,\ldots q_{\rm NM}),$ with $\mathrm{NM}=3*N_A-6$. The corresponding harmonic frequencies are designated by $(\nu_1,\nu_2,\nu_3,\ldots \nu_{\rm NM})$. 
The model is based on the Watson Hamiltonian \cite{Watson1968} with zero rotational angular momentum (J=0).
Its vibrational part contains the PES that is a multidimensional function known only for few points calculated by a first electronic resolution {most commonly accompanied by a chained evaluation of the derivatives\cite{ALLEN1996,Allen1993}. A natural choice of interpolation points would be through the Gauss quadrature rules enhancing polynomial approximation. We can also note the relevance of the ones selected by the Adaptive Density-Guided Approach \cite{Sparta2009,Sparta2010} smoothing the PES where the variations of the energy is the most important.
After this first step involving ab-initio calculations, a fittings is generally performed with classical least square methods \cite{Carter2001,Carter2002} recently improved with Kronecker factorisation \cite{Ziegler2016}.}
{The current version of the code accepts the multivariate polynomial} 
\begin{equation}\label{eqn:poly}
\mathcal{U_K}(\vect{q})=\sum_{ ||\vect{c}||_1\leq \mathrm{DP} } K_{\vect c}\prod_{n=1}^{\mathrm{NM}}q_n^{c_n},
\end{equation}
{identified with}
combinations of monomial degrees $\vect{c}\in \mathbb{N}^{\mathrm{NM}}$ {defined up to a maximal} degree DP and attributed to force constants $K_{\vect c}$. {From a general point of view, the construction fully exploits the  sum of product separability of the PES which makes it a minimal requirement.}
The vibrational part {of the Hamiltonian} writes \begin{equation}\label{eqn:HVib}
\displaystyle\mathcal{H}_{vib}(\vect{q})= -\frac{1}{2}\sum_{n=1}^{\rm NM} \nu_n\frac{\partial^2}{\partial q_n^2} +  \mathcal{U_K}(\vect{q}).
\end{equation}
{The left over coupling terms} consists on the \textit{Coriolis corrections}
\begin{equation}\label{eqn:HRot}
\begin{array}{lcl}
\mathcal{H}_{{CC}}(\vect{q})& = & \displaystyle \mathcal{C}_{ijkl}(\vect{q})-\frac{1}{8}\sum _{\alpha =1}^{3}\mu _{\alpha \alpha },\\
\displaystyle \mathcal{C}_{ijkl}(\vect{q})&=& \displaystyle -\frac{1}{2}\sum_{i<j}\sum_{k<l}Z_{ijkl}\left(\sqrt{\frac{\nu_j}{\nu_i}} q_i\frac{\partial}{\partial q_j}- \sqrt{\frac{\nu_i}{\nu_j}} q_j\frac{\partial}{\partial q_i}\right)\left(\sqrt{\frac{\nu_l}{\nu_k}} q_k\frac{\partial}{\partial q_l}-\sqrt{\frac{\nu_k}{\nu_l}}q_l\frac{\partial}{\partial q_k}\right)\\& & \displaystyle \\ Z_{ijkl}& = & \displaystyle \sum_{(\alpha ,\beta)\in (x,y,z)} \mu_{\alpha,\beta}\zeta_{ij}^{\alpha}\zeta_{kl}^{\beta},
\end{array}
\end{equation}
where $\mu_{\alpha,\beta}$ is the inverse of the moment of inertia {simplified by its constant values obtained at equilibrium geometry} and coefficients $(\zeta_{ij}^{\alpha}, \zeta_{kl}^{\beta})$ are calculated according to the method of Meal and Polo \cite{Meal1956}.
The \textit{Watson term} $-\frac{1}{8}\sum _{\alpha =1}^{3}\mu _{\alpha \alpha }$ {does not modify the transition energies, then it can be independently evaluated and added to the ground state at the end of a vibrational treatment.} 
In regard to the wave function of the total Hamiltonian $\mathcal{H}$, it is a linear combination of basis set elements belonging to an Ansatz $B$
\begin{equation}\label{eqn:phi0}
\begin{array}{lcl}
\displaystyle \Psi(\vect{q}) & = & \displaystyle\sum_{\vect{b} \in B}x_{\vect{b}} \Phi_{\vect{b}}(\vect{q}),\\
\end{array}
\end{equation}
each one writing as a product of one dimensional harmonic {oscillators}
\begin{equation} 
\Phi_{\vect{b}}(\vect{q}) =  \phi_{b_1}(q_{1})\ldots  \phi_{b_n}(q_{n})\ldots  \phi_{b_{\rm NM}}(q_{\rm NM}),\end{equation}
solution of the equation
\begin{equation}
\label{eqn:Harmonic}
\begin{array}{lcl}
\displaystyle\mathcal{H}_0(\vect{q}) \Phi_{\vect{b}}(\vect{q})&=& \displaystyle E_{\vect{b}}\Phi_{\vect{b}}(\vect{q}), \\ \;\;\;\;  \;\;\;\;  \displaystyle E_{\vect{b}}&=& \displaystyle \sum_{n=1}^{\rm NM}\nu_n (b_n+1/2),\\
\displaystyle  \;\;\;\;  \;\;\;\;  \mathcal{H}_0(\vect{q}) &=& \displaystyle\sum_{n=1}^{\rm NM} \frac{\nu_n}{2} \left(-\frac{\partial^2}{\partial q_n^2} + q_n^2\right).\\\end{array} \end{equation}
Implicitly, $\Phi_{\vect{b}}(\vect{q})$ is assimilated to the multi-index ${\vect{b}}$ (cf figure \ref{Fig:KInd}),
and recognized by an integer coinciding with a pointer address when stored in memory. 
The algorithm could also possibly work optimized basis set \cite{Bowman1979,Bowman+1986}, but the efficiency will be impoverished because DVCI fully uses harmonic oscillator properties. 
Afterwards, classical variational formulation leads to the eigenvalue problem
\begin{equation}\label{eqn:HxEx}
H\vect{X}=E \vect{X},
\end{equation}  
with matrix coefficients of $H$ built from the integrals\footnote{Cf Appendix.}
\begin{equation}\label{eqn:VCIELEM}
\begin{array}{lclcl}
\displaystyle  {H_{\vect{s,b}}} & = & \braket{\Phi_{\vect{s}} | \mathcal{H} |\Phi_{\vect{b}}} &=&\displaystyle \int_{\reel^{\mathrm{NM}}}\Phi_{\vect{s}}(\vect{q})\left(\mathcal{H}_{vib}+\mathcal{H}_{{CC}}\right)(\vect{q}) \Phi_{\vect{b}}(\vect{q})\,\\
\\
 \displaystyle \forall (\vect{s},\vect{b}) \in B\times B. & & \\ \end{array} \end{equation}  
\paragraph{The curse of dimensionality}
The number of basis functions exponentially increases with the number of nucleus when adopting a brute force variational method. As example, for a 12-d normal coordinate system, the dimension of the configuration space will be $11^{12}$ for a maximal quantum level equal to 10 in each direction.  To solve the associated eigenvalue problem, one needs to manipulate items with the same size, and a multiplication by 8 gives about 251 terabytes of memory for a double precision vector. 
\section{State of the Art}
In the field of basis selection techniques, one widely uses combination of VCI and perturbation criteria with variation-perturbation theory \cite{Rauhut2007,Scribano2008,Neff2009,Rauhut2008,Baraille2001,Carbonniere2010,
Pouchan1997,Begue2007}. For an member $\Phi_{\vect{b}}$ of a growing subspace $B$, selected basis functions $\Phi_{\vect{s}}$ should verify the perturbation criterion
\begin{equation}\label{eqn:Perturb2} 
\left \vert \frac{\langle \Phi_{\vect{s}}|\mathcal{H}-\mathcal{H}_{0}|\Phi_{\vect{b}} \rangle^2}{E_{\vect{b}} - E_{\vect{s}}} \right\vert > \varepsilon_{\textrm{VP}}, 
 \end{equation}
where $\varepsilon_{\textrm{VP}}$ is a given threshold depending on the accuracy one wants to reach.  $(E_{\vect{s}},\Phi_{\vect{s}})$ and $(E_{\vect{b}},\Phi_{\vect{b}})$ are eigen-pairs of $\mathcal{H}_0$  relying upon the nature of the basis functions that are used. It typically designs the harmonic part \eqref{eqn:Harmonic} or the sum of single-mode VSCF operators \cite{Chaban2000,Chaban1999,Jung1996,Roy2013}.
{A practical manner to increase the configuration space is described in MULTIMODE\cite{MULTIMODE} package, using four classes of excitation namely simple (S), double (D), triple (T) and quadruple (Q). The construction is flexible though it is difficult to guess the optimal combination of excitations that would provide the variational limit.}


To significantly reduce memory usage, matrix entries may not be stocked and computed with pruning conditions \cite{CARNUM2,Avila2011,Avila2011a}. It consists in finding proper weights function $\alpha_n$ calculated on harmonic frequency criteria, and a maximal quantum level $d$ to define the VCI space
\begin{equation}\label{eqn:VCINU}\mathrm{VCI}_{\boldsymbol{\alpha}}(d)=\{\vect{b} \in \mathbb{N}^{\rm NM}, \ \sum_{n=1}^{\rm NM} \alpha_n (b_n)\leq d\}.\end{equation}
For example $\alpha_n (b_n)=\lfloor \nu_n/\nu_{min}+0.5\rfloor * b_n$ \cite{Brown2016}.
When possible, symmetry properties may also be employed to separate basis set into groups of functions belonging to different irreducible representations. 
\hskip 0.2 cm Prospering works on tensorial factorisation relying upon {ALS minimisation \cite{Beylkin2005}} permit a drastic reduction of RAM and time expenditure. In the present context, we can notice the efficacy of the Hierarchical Reduced-Rank Block Power Method (HRRBPM) \cite{Thomas2016,Thomas2015} and the tensor train factorisation\cite{Rakhuba2016}. {As for the MCTDH \cite{Meyer2012,Cao}, the efficiency directly depends on the number of summed products in the PES} that should be small enough to observe accurate results with a low computational cost.\newline
{Among the non variational methods, the VCC theory \cite{Christiansen2004} constitutes a robust way to get precision out of small spaces with a computational cost sharply increasing with the level of excitations exponentially deployed. This effect has been recently mitigated by incorporating the ALS techniques inside the algorithm \cite{Godtliebsen2015,Madsen2017}.}\newline
An other {pertinent} criteria for subspace selection is the residue. It is widely adopted in Davidson like methods \cite{Davidson1975,Ribeiro2005,Ribeiro2002,Zhou2007} and has recently been implemented in A-VCI \cite{Garnier2016,Odunlami2017} sharing same structure than the $A_k$ decomposition \cite{AK}. 
In the $A_k$ theory, one considers primary and secondary spaces respectively called $B$ and $B_{\smalltt{S}}$. In the whole space $B\oplus B_{\smalltt{S}}$, Hamiltonian matrix writes:
\begin{equation}\label{eqn:block}
 H= \left(
\begin{array}{c|c}      
  \\   
 \ \   H_{\smalltt{B}}^{ \ } \ \ &  \qquad  H_{\smalltt{SB}}^{^{T}}  \qquad  \qquad  \\  
    \\    
  \hline
   &     \\
  \\
\ \ H_{\smalltt{SB}} \ \ & H_{\smalltt{S}}  \ \qquad \\
 \\
   &   \\
\end{array}\right),\end{equation}
where the different sub-blocks combine $B$ and $B_{\smalltt{S}}$ in the following way
\begin{equation}\label{eqn:HRRJ}
\begin{array}{lcl}
\displaystyle\left[H_{\smalltt{B}}\right]&=&\displaystyle\left[\braket{\Phi_{\vect{s}} | \mathcal{H} |\Phi_{\vect{b}}}\right]_{(\vect{s},\vect{b}) \in B\times B},\\
\displaystyle\left[H_{\smalltt{SB}}\right]&=&\displaystyle\left[\braket{\Phi_{\vect{s}} | \mathcal{H} |\Phi_{\vect{b}}}\right]_{(\vect{s},\vect{b}) \in B_{\smalltt{S}}\times B},\\
\displaystyle\left[H_{\smalltt{S}}\right]&=&\displaystyle\left[\braket{\Phi_{\vect{s}} | \mathcal{H} |\Phi_{\vect{b}}}\right]_{(\vect{s},\vect{b}) \in B_{\smalltt{S}}\times B_{\smalltt{S}}}.\\
\end{array}
\end{equation}
{Regarding} the A-VCI $B_{\smalltt{S}}$ is included in $\mathcal{H}_{vib}(B)\setminus B$, {namely} the complement of $B$ in its image by $\mathcal{H}_{vib}$\eqref{eqn:HVib}, VMRCI {builds} $B_{\smalltt{S}}\subset\mathrm{STDQ}(B)\setminus B$ and for {PyVCI\_VPT2 $B_{\smalltt{S}}$ is comprised in the $d$-level excitation space written}
\begin{equation}
\label{eqn:refBd} 
{\mathrm{VCI}(d)=\{ \Phi_{\vect{b}} \; / \; |\vect{b}|=\sum_{i=1}^{\mathrm{NM}} b_i \leq d \}.}
\end{equation} 
One can remark the use of the inclusion instead of equality because additional truncation on secondary space may be added to diminish the memory usage.
For example one can cut down the maximal Harmonic energy as for the A-VCI, or only consider STD excitations in the case of VMRCI. The A-VCI is dedicated to compute the first eigenvalues of the Hamiltonian whereas VMRCI and {PyVCI\_VPT2 are focussed on the fundamentals and its degenerates generally constituting the states of interest.} 
For an eigenpair $(E,\vect{X})$ of $H_{\smalltt{B}}$ \eqref{eqn:block}, and $\tilde{\vect{X}}=(\vect{X},\vect{0}_{B_{\smalltt{S}}})$ the zero padded array of $\vect{X}$ in $B\oplus B_{\smalltt{S}}$, the residual vector {and its components write}
\begin{multline}\label{eqn:Rezidu}H\tilde{\vect{X}}-E\tilde{\vect{X}}=(\vect{0}_{B},H_{\smalltt{SB}}\vect{X})^T, \\ {\left(H_{\smalltt{SB}}\vect{X}\right)_{\vect{s}}=\sum_{\vect{b}\in B}H_{\vect{s,b}}x_{\vect{b}}=\braket{\Phi_{ \vect{s} } | \mathcal{H} | \Psi  }, \ \forall \vect{s}\in B_{\mathtt{S}}}.\\\end{multline}
{They measure an error on the energy $E$ and on the wave function $\Psi$ \eqref{eqn:phi0} respectively related to the second and one order corrections expressed in the $A_k$ approximation as}
\begin{multline}\label{eqn:ReziduD}{\Delta E=\sum_{\vect{s}\in B_{\smalltt{S}} }\Delta E_{\vect{s}}=\sum_{\vect{s}\in B_{\smalltt{S}} }\frac{\left(H_{\smalltt{SB}}\vect{X}\right)_{\vect{s}}^2}{E-H_{ \vect{s},\vect{s} } },} \\
 {\Delta \Psi=\sum_{\vect{s}\in B_{\smalltt{S}} }\frac{\left(H_{\smalltt{SB}}\vect{X}\right)_{\vect{s}}}{E+\Delta E-H_{ \vect{s},\vect{s} } } \Phi_{ \vect{s} } .}\\  
 \end{multline}
In an iterative process, A-VCI considers maximal components of {the residual vector whereas VMRCI and PyVCI\_VPT2\footnote{{$H_{ \vect{s},\vect{s} }=\braket{\Phi_{ \vect{s} } | \mathcal{H} |\Phi_{ \vect{s} } }=\langle \Phi_{\vect{s}}|\mathcal{H}-\mathcal{H}_{0}|\Phi_{\vect{s}} \rangle + E_{\vect{s}}$ and only $E_{\vect{s}}$ is retained formula \eqref{eqn:ReziduD} with PyVCI\_VPT2.}} select the configurations from the partial energies $|\Delta E_{\vect{s}}|$ \eqref{eqn:ReziduD} lying above a given threshold impacting the quality of the final results.}
{Whether we consider the residual vector or the correction energy, the selected configurations} are always assigned to {the} secondary space, {then the corresponding errors} will be nullified at next iteration. The expected effect is to minimize the {deviations} between eigenpairs computed in $B$ and the ones that we would have calculated in $B\oplus B_{\smalltt{S}}$. {As a consequence, $B_{\smalltt{S}}$ should be big enough or carefully enlarged to guaranty the pertinence of the measured precision and so the quality of the eigenpairs.} 
 {In this framework of study, there is then trade off between the backing up of the entries of $H_{\smalltt{SB}}$ that would increase the memory requirement and their computation on the fly which might be time consuming.}
\section{Theory and algorithm}\label{NewAlgo}
In this part, we will see that the general algorithm is made from an enhanced mix of the previously recalled methods and a new theoretical approach to the construction of the Hamiltonian.
Here are the main features:
\begin{itemize}
\item For the generation of $B_{\smalltt{S}}$, in addition to the usual ways of expansion previously reminded, the method is capable to calibrate an optimal choice of excitations according to the analysis of the force field and {Coriolis} terms.
\item The precision is focused on a given choice of targets, which allows an effective compression of the information and a significant gain of memory.
\item The presented factorisation relieves the occupation of memory by performing on-the-fly operations for the MVP $H_{\smalltt{SB}}\vect{X}$ \eqref{eqn:Rezidu} that are not compensated by a systematic augmentation 
{of the latency.}
\end{itemize}
\subsection{The local factorisation}
{In the field of mathematics, the duality is a principal that associates two different sets belonging to a same or a distinct structure. A famous example is given by the Riesz representation theorem\cite{rudin1966real} that assimilates a vector space to a set of linear forms (e.g scalar product). In the same order of ideas, we associate a product of creation and annihilation operators with an occupation-number vector \cite{Leaf1973}. This correspondence is based on the ascertainment that the space of occupation numbers can be generated by applying raising and lowering excitations repeatedly on any element of the same space.}
Considering by $\hat{a}_n^{+}$ and $\hat{a}_n^-$ the creation (or raising) and annihilation (or lowering) operators, acting on Hermite functions in the following manner {\cite{cohen1977quantum}}
\begin{equation}\begin{aligned}\hat{a}_n^{+}\ket{\phi_{b_n}}&=\sqrt{b_n+1}\ket{\phi_{b_n+1}},\\
\hat{a}_n^{-}\ket{\phi_{b_n}}&=\sqrt{b_n}\ket{\phi_{b_n-1}},\end{aligned}
\end{equation} 
the position and derivative write
\begin{equation}
\label{eqn:Op2}
\begin{aligned}q_n&=\frac{1}{\sqrt{2}}(\hat{a}_n^{-}+\hat{a}_n^+),\\ \frac{\partial}{\partial q_n} &=\frac{1}{\sqrt{2}}(\hat{a}_n^{-}-\hat{a}_n^+)~.\end{aligned}
\end{equation}
{General second quantized Hamiltonian expressions can be found in \cite{Christiansen2004,Christiansen2007,Cao} while here is used an explicit polynomial representation \cite{Hirata2014,Baiardi2017} noted}
\begin{equation}\label{eqn:SecondH}
{\begin{array}{lcl}
\mathcal{\hat{H}} & = & \displaystyle \sum_{n=1}^{\rm NM} \nu_n\left( \hat{a}_n^{+}\hat{a}_n^{-}+\frac{1}{2} \right)\\
&  + & \displaystyle \sum_{ ||\vect{c}||_1\leq \mathrm{DP} } K_{\vect c}\prod_{n=1}^{\mathrm{NM}}(\hat{a}_n^{-}+\hat{a}_n^+)^{c_n} \\
&  + & \displaystyle \displaystyle -\frac{1}{2}\sum_{i<j}\sum_{k<l}Z_{ijkl}\left(\sqrt{\frac{\nu_j}{\nu_i}}(\hat{a}_i^{-}+\hat{a}_i^+)(\hat{a}_j^{-}-\hat{a}_j^+)- \sqrt{\frac{\nu_i}{\nu_j}} (\hat{a}_j^{-}+\hat{a}_j^+)(\hat{a}_i^{-}-\hat{a}_i^+)\right)\\
& \times & \displaystyle \left(\sqrt{\frac{\nu_l}{\nu_k}} (\hat{a}_k^{-}+\hat{a}_k^+)(\hat{a}_l^{-}-\hat{a}_l^+)-\sqrt{\frac{\nu_k}{\nu_l}} (\hat{a}_l^{-}+\hat{a}_l^+)(\hat{a}_k^{-}-\hat{a}_k^+)\right)\\
\end{array}}
\end{equation}

The local factorisation intends to answer the question:\newline
{Is it possible to develop and factorise expression \eqref{eqn:SecondH} as a sum of product of excitations?
Since the multiplication $\hat{a}_n^{+}\hat{a}_n^{-}$ is not commutative, there is no straightforward response.}


We can, in fact, treat it from another angle by noticing from elementary properties of Hermite functions \footnote{Cf Appendix.} that the paired elements $\braket{\Phi_{\vect{s}} | \mathcal{H}_{vib} |\Phi_{\vect{b}}}\neq 0$ involve only the local force constants\footnote{$K_{\vect c}\neq \nu_n/2$ because it is included in the harmonic part \eqref{eqn:Harmonic} and will be accounted when $\vect{s}=\vect{b}$.} 
\begin{equation}\label{eqn:LFK} \mathrm{LFK}(\vect{s}-\vect{b})=\left\{
\begin{array}{l}\vect{c}\in \mathbb{N}^{\mathrm{NM}},  K_{\vect c}\neq 0,\\
\forall n\in\{1,\ldots,\mathrm{NM}\}, \ K_{\vect c}\neq \nu_n/2,\\
\exists t_n\in \mathbb{N}, \ |s_n-b_n|=c_n-2t_n \end{array} \right\}. \end{equation}
In the same way, for a group of four canonical vectors $(\vect{1}_i,\vect{1}_j,\vect{1}_k,\vect{1}_l)$ of lengths NM with entry 1 respectively on position $i,j,k,l,$ the local \textit{Coriolis} interactions $\braket{\Phi_{\vect{s}} | \mathcal{H}_{{CC}} |\Phi_{\vect{b}}}\neq 0$ contain the combinations
\begin{equation}\label{eqn:LCI}
\mathrm{LCI}(\vect{s}-\vect{b})=\left\{
\begin{array}{l}\vect{c}^{ijkl}=\vect{1}_i+\vect{1}_j+\vect{1}_k+\vect{1}_l ,\\
(i,j,k,l)\in \{1,\ldots ,\mathrm{NM}\}^4 , \\
 i<j, \ k<l, \forall n\in\{1,\ldots,\mathrm{NM}\},\\
\exists t_n\in \mathbb{N}, \ |s_n-b_n|=c_n^{ijkl}-2t_n \\ \end{array} \right\}.
\end{equation}
These sets are gathered in the local force field written
\begin{equation}\label{eqn:LFF}\mathrm{LFF}(\vect{s}-\vect{b})=\mathrm{LFK}(\vect{s}-\vect{b})\cup \mathrm{LCI}(\vect{s}-\vect{b}),\end{equation}
and defined no matter the sign of the differences $(s_n-b_n), \ n\in\{1,\ldots \mathrm{NM}\}$.

With definitions \eqref{eqn:LFK},\eqref{eqn:LCI}, the program builds the set of {occupation-numbers} associated to the non void local force fields\footnote{The zero excitation is systematically included.}
\begin{equation}\label{eqn:DualLFF}\mathrm{LFF}^*=\left\{ \vect{e} \in \mathbb{N}^{\mathrm{NM}}, \ \mathrm{LFF}(\vect{e})\neq\emptyset\right\}\cup\{\vect{0}_{\mathrm{NM}}\},\end{equation} 
{
as depicted in the following loop
\begin{center}
\begin{algorithm}[H]
 \ForAll{$\displaystyle\{(\vect{c},\vect{c}^{ijkl}), \ K_{\vect c}\neq 0, \ i<j, \ k<l\}, \ (i,j,k,l)\in \{1,\ldots ,\mathrm{NM}\}^4$}{~\newline 
  \If{$\displaystyle\forall n\in \{1,\ldots \mathrm{NM}\}, \ \exists t_n\in \mathbb{N}, \ c_n-2t_n \geq 0$}{~\newline~\newline 
   Add $\vect{e}=(c_1-2t_1,\ldots , c_{_{\mathrm{NM}}}-2t_{_{\mathrm{NM}}})$ to LFF$^*$, and $K_{\vect c}$ to LFF($\vect{e}$)~\newline 
   }
  \If{$\displaystyle\forall n\in \{1,\ldots \mathrm{NM}\}, \ \exists t_n\in \mathbb{N}, \ c_n^{ijkl}-2t_n \geq 0$}{~\newline~\newline 
 Add $\vect{e}=(c_1^{ijkl}-2t_1,\ldots , c_{_{\mathrm{NM}}}^{ijkl}-2t_{_{\mathrm{NM}}})$ to LFF$^*$, and $(i,j,k,l)$ to LFF($\vect{e}$)~\newline 
  }
 }
\caption{Building of the set $\mathrm{LFF}^*$. All the force constants and monomial degrees included in a local force field $\mathrm{LFF}(\vect{e})$ are factorised against the excitation $\vect{e}\in \mathrm{LFF}^*$.} 
\label{Algo:BuildLFF}
\end{algorithm}
\end{center}
}

From expressions \eqref{eqn:Op2}, a creation is always accompanied by annihilation. Then, the dual of {$\mathcal{\hat{H}}$ \eqref{eqn:SecondH}} consisting in its intrinsic sum of product of excitations {written as if they were commutative}, is constructed from the positive and negative multi-increments
\begin{equation}\label{eqn:PMLFF}\pm\mathrm{LFF}^*=\left\{(\pm e_1,\ldots,\pm e_{\mathrm{NM}})\in\mathbb{Z}^{\mathrm{NM}}, \ (e_1,\ldots,e_{\mathrm{NM}})\in\mathrm{LFF}^*\right\},\end{equation}
 and {expresses as}\footnote{$\pm e_n$ is abusively employed to indicate the presence of a $+$ or $-$ sign in front of $e_n$.}
\begin{equation}\label{eqn:DualH}\mathcal{H}^*=\sum_{\vect{e}\in \pm\mathrm{LFF}^*}\prod_{n=1}^{\mathrm{NM}}\hat{a}_n^{\pm e_n}.\end{equation}
The spaces $B,B_{\smalltt{S}}$ are growing up together along the repetitions of the main loop. {The set $A$ designs} the added basis functions in $B$ from one iteration to the other. $B_{\smalltt{S}}$ is completed by browsing the image $\mathcal{H}^*(A)$. In the meantime the MVP $H_{\smalltt{SB}}\vect{X}$ is partially calculated for the couples $(\vect{b},\vect{b}+\vect{e})\in (A\times \mathcal{H}^*(A)\setminus B)$ as \footnote{$\mathcal{C}_{ijkl}$ is defined equation \eqref{eqn:HRot}.}
\begin{equation}\label{eqn:MVPPart} \begin{array}{lcl}
\displaystyle (\widetilde{H_{\smalltt{SB}}\vect{X}})_{\vect{b}+\vect{e}}&=&\displaystyle\sum_{\vect{b}\in A}\left[\sum_{\vect{c}\in\{\mathrm{LFK}(\vect{e})\}} K_{\vect{c}}\prod_{n=1}^{\mathrm{NM}}\braket{\phi_{b_n+e_n}|q_{n}^{c_n}|\phi_{b_{n}}}\right.\\
& & \displaystyle  +\left.\sum_{(i,j,k,l)\in \{\mathrm{LCI}(\vect{e})\}}\braket{\Phi_{\vect{b}+\vect{e}}|\mathcal{C}_{ijkl}|\Phi_{\vect{b}}}\right]x_{\vect b}.\\
 \end{array} \end{equation}
The other part of the sum $(\vect{b}\in B\setminus A)$ and other components are then completed as explained in the next section. When setting the parameter DoGraph\ref{DoGraph} to zero, one also has the possibility to directly compute the whole vector $H_{\smalltt{SB}}\vect{X}$ by fetching $\mathcal{H}^*(B)$ instead of $\mathcal{H}^*(A)$ in \eqref{eqn:MVPPart}. {Under these circumstances, a supplement of execution time balanced by a smaller usage of RAM will be observed essentially} because many tests are required to locate the addresses of the members of $B_{\smalltt{S}}$. 
 In {addition} to the classical truncations, the excitations of $\mathcal{H}^*$ can be selected with ThrKX\ref{ThrKX}, defining a threshold on the sum of the force constants contained in each local force field. The purpose is to avoid the runaway of $B_{\smalltt{S}}$ size and incorporate only the most contributive basis functions to the residue. 
\subsection{Complementary storage}\label{MatElem}
Solving an eigenvalue problem usually requires a significant amount of MVPs, then the non null coefficients of $H_{\smalltt{B}}$ \eqref{eqn:HRRJ} might rather be collected than evaluated on the fly. 
So far, for any related method, the complexity to determine a non null matrix entry is at least of order $O\left(\mathrm{NM*NPES}\right),$ where NPES designates the total number of force constant in the PES. 
When using the local force field for the evaluation of $\braket{\Phi_{\vect{s}}|\mathcal{H}|\Phi_{\vect{b}}}$, it is enough to fetch \begin{equation}(\vect{s}-\vect{b})^+=(|s_1-b_1|,\ldots , |s_{\mathrm{NM}}-b_{\mathrm{NM}}|)\mathrm{ \ into \ }\mathrm{LFF}^*\eqref{eqn:DualLFF},\end{equation} calculate the terms in bracket{s} equation \eqref{eqn:MVPPart} for $\vect{e}=\vect{s}-\vect{b}$ and add the harmonic energy $E_{\vect{b}}$ when $\vect{b}=\vect{s}$. 
The localization of the position of $(\vect{s}-\vect{b})^+$ in $\mathrm{LFF}^*$ is performed with a binary search \cite{BSEARCH} costing $O\mathrm{(NM*log(|\mathrm{LFF}^*|))}$ operations. Consequently, the total complexity is \begin{equation}\label{eqn:cost}O\left(\mathrm{NM}*\left[|\mathrm{LFF}(\vect{s}-\vect{b})|+\mathrm{log(|\mathrm{LFF}^*|)}\right]\right),\end{equation} and one can easily check that $|\mathrm{LFF}(\vect{s}-\vect{b})|+\mathrm{log(|\mathrm{LFF}^*|)}<<\mathrm{NPES}$. Indeed for $\vect{e}\neq \vect{0}_{\mathrm{NM}}$\footnote{The case $\vect{e}=\vect{0}_{\mathrm{NM}}$ intervenes only for the construction of the diagonal elements of $H_{\smalltt{B}}$.}, the cardinal of $\mathrm{LFF}(\vect{e})$\eqref{eqn:LFF} is always very much lower than NPES. The worst case being NPES/2*NM happening only for a PES with no crossings.
It therefore appears that the local factorisation consumes less operations than traditional methods for the determination of matrix coefficients, and for a small amount of MVPs, will be capable to integrate into a calculation on the fly without jeopardizing the execution time. Nevertheless, it will be all the more accelerated as the address of the paired elements is known in advance.
This technique is actually employed for the residual block when the parameter DoGraph is strictly positif, by keeping only the pointers on the non zeros of $H_{\smalltt{SB}}$, thus directly accessible to complete the MVP \eqref{eqn:MVPPart} for the missing parts: \begin{equation}\label{eqn:Complet}(H_{\smalltt{SB}}\vect{X})_{\vect{s}}=(\widetilde{H_{\smalltt{SB}}\vect{X}})_{\vect{s}}+\sum_{\vect{b}\in (B\setminus A)}\braket{\Phi_{\vect{s}}|\mathcal{H}|\Phi_{\vect{b}}}x_{\vect{b}}, \ \forall\vect{s}\in \mathcal{H}^*(B)\setminus B.\end{equation}
\subsection{Initial space construction} 
  Let's consider the ordered eigenvalues of block $H_{\smalltt{B}}$ built at iteration $\mathtt{i}$ \begin{equation}\label{eqn:num} E_{0}^{\mathtt{i}}\leq \ldots \leq E_{\ell}^{\mathtt{i}} \leq E_{\ell+1}^{\mathtt{i}}\leq \ldots. \end{equation} 
One can demonstrate with the Poincar\'e separation theorem\cite{Poincare} that each $E_{\ell}^{\mathtt{i}}$ is a decreasing sequence of $\mathtt{i}$. Consequently, the minimization of the differences
\begin{equation} E^{\mathtt{i}}_{\ell}-E_{\ell}^{\mathtt{i}+1} \end{equation}
could be effective only if no spectral hole is introduced in the interval holding targets at step zero where the eigenvalues are simple harmonic energies \eqref{eqn:Harmonic}. Also, in order to integrate the perturbation effects, the initial space writes
\begin{equation}\label{eqn:B0} B= \{\vect{b} \ / \ E_{\vect{b}} \leq E_{max}*\kappa\},
\end{equation} 
where \footnote{$\displaystyle E_{\vect{0}}=\frac{1}{2}\sum_{n=1}^{\mathrm{NM}} \nu_n$.}$E_{max}=\mathrm{MaxFreq}+E_{\vect{0}}$, MaxFreq\ref{MinFreq} is the maximal tracked frequency and $\kappa$ \ref{kappa} an empirical elongation factor accounting the global deviation between converged eigenvalues and  associated harmonic energies. Its default value is set to 1.2, but it is automatically increased in agreement with the anharmonicity growing up as we get away from the smallest eigenvalue. 
To avoid timing issues caused by the great number of combination band\footnote{$\displaystyle\sum_n^{\mathrm{NM}} b_n\nu_{n}$.} to be tested, the initial space is calculated by recursively applying {simple raising excitations starting from the zero configuration until size consistency.}

\section{The iterative process}
The maximal size of the arrays used for $B,B_{\smalltt{S}},H_{\smalltt{B}}$ and $H_{\smalltt{SB}}$ is determined with the allocated memory controlled by the parameter Memory \ref{Memory}.
{The direct sum $B\oplus B_{\smalltt{S}}$ evolves in the product space :}
\begin{equation}\label{eqn:Pid} {\prod_{n=1}^{\mathrm{NM}}[0,\rm d_n],}\end{equation}
{where degrees $d_n$ are defined by Freq0Max\ref{Freq0Max} and MaxQLevel\ref{MaxQLevel} as follows}
\begin{equation}\label{eqn:MaxQ}{d_n=\min(\lfloor \frac{\mathrm{Freq0Max}}{\nu_n}\rfloor, \mathrm{MaxQLevel})}.\end{equation}
{Note that Freq0Max is also the maximal allowed value of harmonic combination bands that is equivalent to the following pruning condition
\begin{equation}\label{eqn:Prun}\sum_{n=1}^{\mathrm{NM}}\left(\frac{\nu_n}{\nu_{min}}\right)*b_n\leq \frac{\mathrm{Freq0Max}}{\nu_{min}}.\end{equation}
Naturally, the resulting reference space is never fully browsed over all the possible configurations, but it
constitutes a barrier for the growth of $B_{\mathtt{S}}$ that is recursively enlarged from $\mathcal{H}^*$ \eqref{eqn:DualH}.}
At any iteration, the set $\{\mathrm{Targ}\}$ refers to eigenvectors of $H_{\smalltt{B}}$ having one component larger than ThrCoor \ref{ThrCoor} and assigned to targets given by the parameter TargetState \ref{Target}. 
After a prior construction of the objects $\left\{\mathrm{LFF}(\vect{e}), \ \vect{e}\in \mathrm{LFF}^*\right\}$\eqref{eqn:LFF}, the sequence of successive main steps decomposes as
\begin{enumerate}[label=$<$\arabic*$>$]
\item Build the initial subspace \eqref{eqn:B0}.
\item\label{Loop}
Compute the eigenpairs $(E_{\ell},\vect{X}_{\ell})$ of $H_{\smalltt{B}}$ with the Implicitly Restarted Lanczos Method of ARPACK \cite{Lanczos1950,ARPACK}. The upper limit of calculated eigenvalues is chosen with MaxEv \ref{MaxEv}. 
\item
Evaluate the residual vectors on the fly $H_{\smalltt{SB}}\vect{X}_{\ell}, \ \ell\in \{\mathrm{Targ}\}$ and secondary basis set $B_{\smalltt{S}}$ as explained in section \ref{NewAlgo}. The expansion of $B_{\smalltt{S}}$ can additionally be reduced with an elimination of the less contributing excitations of $\mathcal{H}^*$ via the parameter ThrKX\ref{ThrKX}. 
Biggest components (in absolute value) of residual vectors are then employed to select basis functions to be added for next iteration through the inputs NAdd \ref{NAdd}, EtaComp \ref{EtaComp} and MaxAdd \ref{MaxAdd}.
\item Go back to step \ref{Loop} as long as {the maximal relative residue stays above a given threshold namely}
\begin{equation}\label{eqn:RelRez}\max_{\ell \in \{\mathrm{Targ}\}}\frac{\Vert H_{\smalltt{SB}}\vect{X}_{\ell}\Vert}{E_{\ell}} > \mathrm{EpsRez} \ \ \text{\ref{EpsRez}}.\end{equation}
This criterion can be fulfilled only if enough memory has been allocated at the beginning. Otherwise, the algorithm will stop until maximal number of basis functions has been reached. It also trivially appears that increasing the targets does also augment the required memory to make them converge at once. 
\end{enumerate}
\section{Benchmarks}\label{Examples}
All the calculus are done on a 64 bits, 2.70GHz quad core processor (model Intel i5-3340M) with 8 Gigabytes of RAM. 
No parallel process is used in here. DVCI program runs on a single CPU so that the actual computational time is the same as the CPU wall time. {The memory unit is the Megaoctet (MO)equivalently known as the Megabyte (MB). On reminder, the relative residues\eqref{eqn:RelRez}, and correction energies $\Delta E$ \eqref{eqn:ReziduD} are calculated in the secondary space $B_{\mathtt{S}}=\mathcal{H}^*(B)\setminus B$ truncated with the pruning condition (\ref{eqn:MaxQ}, \ref{eqn:Prun}) and generated with the most contributive excitations of $\mathcal{H}^*$ \eqref{eqn:DualH} selected with ThrKX. For the default parameter values, a calculation carried out to the end with EpsRez $=0.008$ will deviate around $\rm 1cm^{-1}$ from the reference. Another indicator of convergence is given by the height of the eigenvalues which is decreasing according to the Poincar\'e separation theorem.
Shrinking the reference space by lowering down the value of (Freq0Max, MaxQlevel), will certainly diminish the required computational resources, yet it will be difficult to predict which values will provide the variational limit. It is also possible to raise the threshold for the matrix elements, but for the benchmarks presented here, one tends to avoid cutting down matrices and reference spaces to get more chances to actually converge.
To visualize the pruning condition \eqref{eqn:Prun} with a linear combination of quantum numbers, the weights $\left(\nu_n/\nu_{min}\right)$ are rounded to the closest integers.}
\subsection{$\rm{CH_3CN}$: Acetonitrile}
Methods are compared on molecule $\rm{CH_3CN}$ with the same 
PES as the one used in Ref.~\cite{Avila2011,Leclerc2014,Thomas2015,Garnier2016} 
that was initially introduced by B\'egu\'e and al.~\cite{Begue2005}, computed at CCSDT/cc-pVTZ level 
for harmonic frequencies and B3LYP/cc-pVTZ for higher order terms. 
This PES counts 311 terms (12 quadratic, 108 cubic and 191 quartic).
The benchmark results are taken from Avila et Carrington \cite{Avila2011} where symmetry has been employed to separate the full VCI space into two smaller subspaces. This PES has a small number of non null derivatives regarding the size of the molecule.
{The variational space defined in \cite{Avila2011} is the pruned basis set}
\begin{equation} \left\{\begin{array}{l}\displaystyle \vect{b} \in \mathbb{N}^{12}, \ \sum_{n=1}^{12} \alpha_n b_n \leq 27 ,\\
\displaystyle \alpha_1 = 3, \alpha_2 = 4, \alpha_3 = 3, \alpha_4 = 3, \alpha_5 = \alpha_6 = 3,\\
\displaystyle \alpha_7 = \alpha_8 = 4, \alpha_9 = \alpha_{10} = 3, \alpha_{11} = \alpha_{12} = 1.\end{array} \right\}
\end{equation} 
It contains $743\,103$ harmonic functions and 
the fundamental harmonic frequencies are 
\begin{eqnarray*}
\nu_1&=&3\,065,\,\nu_2=2\,297,\,\nu_3=1\,413,\,\nu_4=920,\nonumber\\
\nu_5&=&\nu_6=3\,149,\,\nu_7=\nu_8=1\,487,\nonumber\\
\,\nu_9&=&\nu_{10}=1\,061,\,\nu_{11}=\nu_{12}=361\,(\text{cm}^{-1}).\nonumber
\end{eqnarray*}~\newline  
{The default values (Freq0Max,MaxQLevel)=(30000 $\mathrm{cm}^{-1}$,15) induce the following rounded pruning condition
\[\left\{\begin{array}{l}\displaystyle \vect{b} \in \mathbb{N}^{12}, \ \vect{b}\leq (9, 13, 15, 15, 9, 9, 15, 15, 15, 15, 15, 15)\\ 
\displaystyle  8b_1 + 6b_2 + 4b_3 + 3b_4 + 9b_5 + 9b_6 + 4b_7 + 4b_8 + 3b_9 \\
 \displaystyle  + 3b_{10} + b_{11} + b_{12} \leq 83\end{array} \right\}.\]
This space is rather huge (712 713 289 elements) and the calculus could be exact in a much smaller one, but it shows that there is almost no limitation on its choice.}  
\paragraph{Results}~\newline
\begin{longtable}{|c|c|c|c|c|c|}
\caption[]{Acetonitrile anharmonic fundamental frequencies, {followed by the relative residues, the correction energies \eqref{eqn:ReziduD}, the absolute errors relative to the reference calculation and the experimental values.}}\\
\hline
 Assignment & Freq & Relative & {$\Delta E$} & Absolute error & Exp   \\
 (Component) & (position)  & Residue &  & {Ref-Here} & values \\ 
\hline
$\nu_{0}$(0.97) &  9837.43(0) &  0.0015 & -0.0171 &  ? & \\ 
$\nu_{11}$(0.97) &  361.11(1) &  0.0034 & -0.1238 &  -0.1198 & 362  \cite{Shimanouchi1977} \\ 
$\nu_{12}$(0.97) &  361.17(2) &  0.0039 & -0.1750 &  -0.1779 & \\ 
$\nu_{4}$(0.95) &  900.76(6) &  0.0032 & -0.1087 &  -0.1001 & 920  \cite{Shimanouchi1977}\\ 
$\nu_{9}$(0.97) &  1034.25(7) &  0.0033 & -0.1228 &  -0.1202 & 1041  \cite{Shimanouchi1977} \\ 
$\nu_{10}$(0.97) &  1034.35(8) &  0.0041 & -0.2133 &  -0.2229 & \\ 
 & & & & &\\
$\nu_{3}$(0.74), $\nu_{9}+\nu_{11}$(0.44),&  1389.17(15) &  0.0038 & -0.1885 & -0.1980 & 1385 \cite{Shimanouchi1977} \\ 
$\nu_{10}+\nu_{12}$(0.44) & & & & & \\ 
 & & & & &\\
$\nu_{3}$(0.62), $\nu_{9}+\nu_{11}$(0.52),&  1397.94(17) &  0.0042 &  -0.2368 &-0.2485 &  1402 \cite{Paso1994}\\ 
$\nu_{10}+\nu_{12}$(0.52)& & & & & \\
 & & & & &\\
$\nu_{7}$(0.97) &  1483.33(20) &  0.0031 & -0.1176 &  -0.1034 & 1450 \cite{Paso1994} \\ 
$\nu_{8}$(0.97) &  1483.46(21) &  0.0040 &  -0.2341 & -0.2331 & \\ 
$\nu_{2}$(0.90) &  2250.94(70) &  0.0037 &  -0.1923 & -0.2157 & 2267  \cite{Shimanouchi1977}\\ 
  & & & & &\\
$\nu_{1}$(0.60), $2\nu_{7}$(0.51), &  2947.42(187) &  0.0043&-0.3469 &  -0.3665 & \\ 
  $2\nu_{8}$(0.51)& & & & &\\
    & & & & &\\
$\nu_{1}$(0.67), $2\nu_{7}$(0.45), &  2981.01(193) &  0.0037& -0.2522 & -0.2316 & 2954 \cite{Shimanouchi1977}\\ 
  $2\nu_{8}$(0.45)& & & & & \\
 & & & & &\\
$\nu_{5}$(0.93) &  3049.12(218) &  0.0039&-0.2796 &  ? & 3009  \cite{Shimanouchi1977}\\ 
$\nu_{6}$(0.93) &  3049.16(219) &  0.0040& -0.3150&  ? & \\ 
\hline
\end{longtable}
\paragraph{Performances summary}~\newline
\begin{longtable}{|c|c|c|c|c|c|}
\caption[]{Performances summary on Acetonitrile fundamental targets. The CPU wall time is in second with the total number of iterations showed in parenthesis. EtaComp=3(cf\ref{EtaComp}), NAdd=50 (cf \ref{NAdd}), {ThrKX$=10^{-15}$}.}\\
\hline
  Final & Final  &   Final  &  Final  &CPU Wall time(s)& Memory \\
  size of $B$ & size of $B_{\smalltt{S}}$ & $nnz(H_{\smalltt{B}})$ & $nnz(H_{\smalltt{SB}})$ & (Iterations) & usage (MO)\\ 
  \hline
6169 &  590840 & 247662 & 3213078 & 352(4) & {115.6} \\
\hline
\end{longtable}
The complexity of this problem {stems} from the fact that the fundamental anharmonic frequencies are dispatched far away from the extremity of the spectrum.
In the recent work of Ondunlami et al\cite{Odunlami2017}, the A-VCI didn't catch up all of them when computing the first 238 eigenvalues.
Their best declared results parallelized on a 24-core Intel Xeon E5-2680 processors running at
2.8 GHz shows a maximal absolute error (relatively to the same reference) equal to 0.305 $\rm{cm}^{-1}$ on the first 121 eigenvalues with a time equal 5637 seconds and a final basis set of $86\,238$ elements.   
In the HRRBPM of Thomas et Carrington\cite{Thomas2015} all the anharmonic frequencies before 2209 $\rm{cm}^{-1}$ were computed with a error lower than 0.38 $\rm{cm}^{-1}$, 3.2 hours cpu time and {115.6} MB on a single Intel Core i7-4770 processor running at 3.4 GHz.
 In here all the fundamentals are computed with {115.6} megabytes memory for a maximal error lower than $0.37~\rm{cm}^{-1}$ and a time of 5 minutes 52 seconds on a single CPU running at 2.70GHz.
\subsection{$\rm{C_2H_4}$ : Ethylene}
The potential energy surface is originally the sixth order curvilinear symmetry-adapted coordinates of Delahaye \& al \cite{Delahaye2014} transformed into the sextic normal coordinate force field with PyPES \cite{PyPES}.
In there work, point symmetry group D2h of $\rm{C_2H_4}$ , has been exploited to divide VCI matrix into 8 symmetry blocks of respective dimension
106 889, 101 265, 100 366, 105 518, 105 643, 101 145, 101 255, 105 697.
No symmetry assumption is applied in here.

{This study is supported by a comparison with the software PyVCI\_VPT2 previously introduced. Although the two methods have some notable distinctions, we try to match different parameters. The thresholds on the matrix elements and force constants are set to $10^{-15}$ in both cases.} 

\paragraph{Harmonic Frequencies and derivative orders of the PES}~\newline
\begin{equation*}
\begin{array}{lll}
 \nu_{1} : 825.0 , &   \nu_{2} : 950.2 , &  \nu_{3} : 966.4 ,  \\ 
 \nu_{4} : 1050.8 , &  \nu_{5} : 1246.8 , &  \nu_{6} : 1369.4 ,  \\ 
 \nu_{7} : 1478.5 , &  \nu_{8} : 1672.6 , &  \nu_{9} :  3140.9 ,   \\ 
 \nu_{10} : 3156.8 , &  \nu_{11} : 3222.9 , &  \nu_{12} : 3248.7 .  \\
 \end{array}
\end{equation*}
\begin{tabular}{c|c|c|c|c|c}
Derivative order & 2 & 3 & 4 & 5  & 6 \\
\hline
& & & & & \\
Number of terms & 45 & 147 & 290 & 642 & 1732 \\
& & & & & \\
\end{tabular}~\newline~\newline
The second derivatives contain the harmonic and non null crossed terms.
{ 
The thresholds for basis state selection have been chosen so that the adjustment on the fundamentals is of the order $1\mathrm{cm}^{-1}$ respectively observable for $\rm VCI\_VPT2\_ETHRESH=10^{-8}$ (cf \href{https://sourceforge.net/projects/pyvci-vpt2/files/docs/PyVCI\_VPT2\_manual.pdf/download}{PyVCI\_VPT2 manual}) and $\rm EpsRez=0.0068$. The reference space for PyVCI is $\mathrm{VCI}(8)$ \eqref{eqn:refBd}, which appears to be quite small ($125\,270$ basis functions), but complete enough to get good accuracy on the fundamentals. A further study shows that this is not exact for all the frequencies close to the mid infrared limit, in particular for $2\nu_1+ \nu_6$ that is strongly coupled with $\nu_{10}$. For DVCI the paired parameters (Freq0Max,MaxQLevel)=(24000 $\mathrm{cm}^{-1}$,10) translate into the following pruned space
\[\left\{\begin{array}{l}\displaystyle \vect{b} \in \mathbb{N}^{12}, \ \vect{b} \leq ( 10, 10, 10, 10, 10, 10, 10, 10, 7, 7, 7, 7 ) \\
\displaystyle b_{1} + b_{2} + b_{3} + b_{4} + 2b_{5} + 2b_{6} + 2b_{7}  +2b_{8} + 4b_{9}  \\
 \displaystyle + 4b_{10} + 4b_{11} + 4b_{12} \leq 29\end{array} \right\},\]
 totalizing 15 896 872 elements.}
\pagebreak 
 \paragraph{Results}~\newline
{\begin{longtable}{|c|c|c|c|c|c|c|c|}
\caption[]{Ethylene anharmonic fundamental frequencies {and the isolated target $2\nu_{1}+\nu_{6}$ calculated with DVCI and PyVCI\_VPT2 completed by the full VCI results in VCI(8). Then appears the second order correction energies \eqref{eqn:ReziduD}. For PyVCI\_VPT2, the computed frequencies in VCI(8) are the same when included in VCI(10).}}\\
\hline
Assign&Freq &Freq &Freq &$\Delta E$&$\Delta E$& $\Delta E$ \\
 (Comp) & DVCI\footnote{{Watson term $-\frac{1}{8}\sum _{\alpha =1}^{3}\mu _{\alpha \alpha }=-1.69$}}&PyVCI &Full VCI &(DVCI)&PyVCI&PyVCI\\ 
    &(Position)&VPT2  & VCI(8)\footnote{{$\rm |VCI(8)|=125270 \ elements$}}  &  & VPT2 & VPT2\\ 
       & & $\rm\subset VCI(8)$ &  &  &$\rm\subset VCI(8)$ & $\rm\subset VCI(10)$\\  
\hline
$\nu_{0}$(0.98) & 11017.09(0)& 11017.26&11016.96 & -0.14 & -0.3 & -0.32\\ 
$\nu_{1}$(0.98) & 823.76(1) & 824.17 &823.66 & -0.35 & -0.48 & -0.6\\ 
$\nu_{2}$(0.98) & 935.48(2) & 935.95 &935.31 &-0.42 & -0.59 & -0.70\\ 
$\nu_{3}$(0.98) & 950.90(3) & 951.38 &950.74 &-0.42 & -0.58 & -0.70\\ 
$\nu_{4}$(0.98) & 1026.09(4) & 1026.52 &1025.92 &-0.41 & -0.55 & -0.65\\ 
$\nu_{5}$(0.98) & 1225.10(5) & 1225.33 &1224.87 &-0.44 & -0.43 & -0.5 \\ 
$\nu_{6}$(0.97) & 1342.94(6) & 1343.15 &1342.85 &-0.29 & -0.27 & -0.32 \\ 
$\nu_{7}$(0.98) & 1442.09(7) & 1442.37&1441.84 &-0.45 &-0.49 & -0.56 \\ 
$\nu_{8}$(0.90) & 1625.55(8) & 1626.10 &1625.41 &-0.41 & -0.63 & -0.75 \\ 
$\nu_{9}$(0.86) & 2986.09(60) & 2986.71 &2985.48 &-0.69 & -1.08 & -1.22 \\ 
$\nu_{10}$(0.80) & 3019.35(62) & 3020.26 &3019.15 &-0.58 & -0.99& -1.25 \\ 
$\nu_{11}$(0.91) & 3080.03(65) & 3080.41 &3079.36 &-0.66 & -0.92 & -1.05 \\ 
$\nu_{12}$(0.93) & 3102.00(70) & 3102.32&3101.26 &-0.58 & -0.88 & -1 \\ 
\hline
&&&&&&\\
$2\nu_{1}+\nu_{6}$ & 3007.39(53) & 3008.59 & 3008.52 &-0.85 & -0.48 & -1.64 \\
(0.84)&&&&&&\\
\hline
\end{longtable}}

\paragraph{Performance summary}~\newline
{\begin{longtable}{|c|c|c|c|c|c|c|}
\caption[]{{DVCI} performances summary on ethylene fundamentals {and target $2\nu_{1}+\nu_{6}$}. The CPU wall time is in second with the total number of iterations indicated in parenthesis. EtaComp=2, {NAdd=(80,200) (for each target), EpsRez=(0.0068,0.0073), ThrKX$=1$}.}\\
\hline
 Targets & Final & Final  &   Final  &  Final  &CPU Wall time& Memory \\
  & size of $B$ & size of $B_{\smalltt{S}}$ & $nnz(H_{\smalltt{B}})$ & $nnz(H_{\smalltt{SB}})$ & (Iterations) & usage (MO)\\ 
  \hline
Fund& 15549 &  1914601 & 4902728  & 40719820
& 20 min 43 s (6) & 378\\
\hline
$2\nu_{1}+\nu_{6}$ &4368 &  555782 & 1195293 & 12283453 
& 3 min (6) & 83\\
\hline
\end{longtable}}
Even if the number of normal coordinates is the same as in the previous system, the memory requirement and the CPU time are significantly increased due to the higher number of terms in the PES.
{The state $2\nu_1+ \nu_6$ has been isolated from the fundamentals with DVCI by setting ThrCoor\ref{ThrCoor} to 0.65 whereas it is automatically included in PyVCI. 
Three series of calculations have been launched for PyVCI. The first one is the full VCI calculation taking about 3 days and 3GO of RAM. The second one with the VPT2 selection in VCI(8) lasted more than 9 hours compensated by a low buffer storage thanks to a clever system of backup of VCI matrix elements in binary files on the hard drive. The maximal memory expend recorded from the command line was 226 MO of RAM and 395 MO of ROM. The third calculation was included in $\rm VCI (10)$ (containing 646646 elements) instead of VCI(8). It has spread over a period of about one day and a half with an utilization of 953 MO of ROM and 653 MO of RAM. Thus we notice that enlarging the reference space absorbs more computational resources and mechanically increases the correction energies especially for the degenerated target $2\nu_1+ \nu_6$ who is shifted by more than $1\mathrm{cm}^{-1}$. From this observation, we deduce that VCI(8) is not large enough to be considered as a reference for the variational limit when the computation is not strictly reduced to the fundamentals. \newline
\hspace*{0.2truecm} In the case of DVCI, the quantum levels are very little restricted, which makes it possible to customize the secondary space and thus more correctly assimilate the correction energy to a real error. Besides, the frequencies have a better accuracy and the latency is reduced by more than a factor 20 while spending less memory even though the reference space is around 126 times larger than VCI(8) and 25 times VCI(10).}  
\subsection{$\rm{C_2H_4O}$ : Ethylene oxide}
Here is computed the fundamentals
of a 15-d Hamiltonian system where the PES is the one of B\'egu\'e and al \cite{Begue2007} calculated
at the CCSD(T)/cc-pVTZ level for harmonic frequencies and B3LYP/6-31+G(d,p) for the other terms. Apart from the harmonic force constants, the PES contains 180 cubic and 445 quartic terms.
The reference results are taken from the A-VCI\cite{Odunlami2017} where final basis set contains 7 118 214 elements.
\paragraph{Harmonic frequency in $\mathrm{cm}^{-1}$}~\newline
\begin{eqnarray*}
\nu_{1} : 3117.9 , \ \nu_{2} : 1549.1 , \ \nu_{3} : 1300.1  ,\ \nu_{4} : 1157.9, \ \nu_{5} : 899.6 , \\ \nu_{6} : 3196.6 , \nu_{7} : 1176.0, \ \nu_{8} : 1052.2 , \ \nu_{9} : 3109.5, \ \nu_{10} : 1512.3 , \\ \nu_{11} : 1156.8 , \ \nu_{12} : 850.2 , \ \nu_{13} : 3211.3 , \ \nu_{14} : 1175.0 , \ \nu_{15} : 815.5 , \\
\end{eqnarray*}
Correspondence with the ones listed in \cite{Odunlami2017} 
\begin{eqnarray*}
\nu_{1} \equiv \omega_{13} , \ \nu_{2} \equiv \omega_{11} , \ \nu_{3}\equiv \omega_9  ,\ \nu_{4} \equiv \omega_6, \ \nu_{5}\equiv \omega_3 , \\ \nu_{6} \equiv \omega_{14} , \nu_{7} \equiv \omega_8, \ \nu_{8} \equiv \omega_4, \ \nu_{9}\equiv \omega_{12}, \ \nu_{10} \equiv \omega_{10} , \\ \nu_{11}\equiv \omega_5 , \ \nu_{12} \equiv \omega_2 , \ \nu_{13} \equiv \omega_{15} , \ \nu_{14} \equiv \omega_7 , \ \nu_{15} \equiv \omega_1 . \\
\end{eqnarray*}~\newline
{The maximal harmonic frequency Freq0Max=30000 $\mathrm{cm}^{-1}$ additionally truncated by MaxQlevel=15, leads to the rounded pruning condition
\[\left\{\begin{array}{l}\displaystyle \vect{b} \in \mathbb{N}^{15}, \ \vect{b}\leq (9, 15, 15, 15, 15, 9, 15, 15, 9, 15, 15, 15, 9, 15, 15)\\ 
\displaystyle  4b_1 + 2b_2 + 2b_3 + b_4 + b_5 + 4b_6 + b_7 + b_8  \\
\displaystyle +4b_9 + 2b_{10} + b_{11} + b_{12} + 4b_{13} + b_{14} + b_{15} \leq 17\end{array} \right\}.\]
}
%
\paragraph{Results}~\newline
\begin{longtable}{|c|c|c|c|c|}
\caption[]{Ethylene oxide anharmonic fundamental frequencies for five groups of targets separated by an horizontal bar. For comparison purpose, the zero point energy has the same value than in \cite{Odunlami2017} (i.e $12461.473 \ \rm{cm}^{-1}$).}\\
\hline
 Assignment & Freq & Relative & {Correction} & Error  \\
 (Component) & {(position)} & Residue & {energy\eqref{eqn:ReziduD}}  & {Ref-Here} \\ 
\hline 
$\nu_{0}$(0.98) & 12461.47(0) & 0.0034 & -0.1285  & -0.1445 \\
$\nu_{15}$(0.97) & 792.96(1) & 0.0045 &  -0.2657 & -0.1850 \\ 
$\nu_{12}$(0.96) & 822.19(2) & 0.0041 & -0.2341 & -0.1347 \\ 
$\nu_{5}$(0.97) & 878.51(3) & 0.0038 & -0.2016 & -0.0895 \\ 
$\nu_{8}$(0.97) & 1017.47(4) & 0.0044 & -0.2722 & -0.1864 \\ 
$\nu_{4}$(0.96) & 1121.47(5) & 0.0042 & -0.2445 & -0.1539 \\ 
$\nu_{11}$(0.97) & 1123.92(6) & 0.0042 & -0.2445 & -0.1507 \\   
$\nu_{14}$(0.97) & 1146.03(7) & 0.0042 & -0.2605 & -0.1622 \\   
$\nu_{7}$(0.97) & 1148.19(8) & 0.0037 & -0.1943 & -0.0845 \\ 
$\nu_{3}$(0.94) & 1271.17(9) & 0.0047 & -0.3303 & -0.2454 \\ 
$\nu_{10}$(0.97) & 1467.58(10) & 0.0037 & -0.2008 & -0.0947 \\  
$\nu_{2}$(0.94) & 1495.49(11) & 0.0042 &  -0.2689 & -0.1879 \\   
\hline
$\nu_{9}$(0.64), $\nu_{2}+\nu_{10}$(0.52) & 2906.77(95) & 0.0041 & -0.2786 & -0.4562 \\ 
$\nu_{9}$(0.52), $\nu_{2}+\nu_{10}$(0.63) & 2989.70(111) & 0.0044 & -0.3266 & -0.4552 \\ 
\hline
$\nu_{1}$(0.45), $2\nu_{10}$(0.62), & 2916.94(99) & 0.0040 &-0.2328 & -0.2640 \\ 
 $\nu_{8}+\nu_{11}+\nu_{15}$(0.44) & & & &\\
..&..&..&..&..\\
$\nu_{1}$(0.52), $2\nu_{10}$(0.62) & 2952.86(103) & 0.0047 & -0.4016 & -0.4654 \\ 
\hline
$\nu_{6}$(0.85) & 3025.71(116) & 0.0036 & -0.2396 & -0.3801 \\
\hline
{$\nu_{13}$(0.80)} & {3037.31(118)} & {0.0038} & {-0.2034} & {-0.3955} \\ 
\hline
\end{longtable}
\paragraph{Performances summary}~\newline
\begin{longtable}{|c|c|c|c|c|c|c|}
\caption[]{Performances summary on ethylene oxide. Each tracked state(s) are separated by an horizontal bar. The CPU wall time is in second with the total number of iterations showed in parenthesis.
NAdd=100 for all targets except for $\nu_{13}$ and $\nu_6$ where NAdd=300. EtaComp=3 and {ThrKX$=10^{-15}$} in any case.}\\
\hline
Target(s) & Final & Final  &   Final  &  Final  & Wall time(s)& Memory \\
 & size of $B$ & size of $B_{\smalltt{S}}$ & $nnz(H_{\smalltt{B}})$ & $nnz(H_{\smalltt{SB}})$ & (Iterations) & usage (MO)\\ 
\hline
$\nu_{15},\nu_{12}$ & & & & &  &\\
$\nu_{5},\nu_{8},$ & & & & & & \\
$\nu_{4},\nu_{11},$ & 83346 &  11182617 &  5396776 &  80959806 & 11823(7) & {1197.3} \\
$\nu_{14},\nu_{7},$ & & & & & & \\
$\nu_{3},\nu_{10}$ & & & & & & \\
\hline
$\nu_{9}$ & 133128 & 11840777 & 13352241 & 126500214 & 21507(16) & {1383.6} \\ 
\hline
$\nu_{1}$ & 121180  & 11788217 & 11420147 & 117896340 & 21430 (16) & {1347.2} \\ 
\hline
$\nu_{6}$ & 84554  & 8374499 & 7341322 & 79908192 & 11299(13)  & {922.5} \\
\hline
$\nu_{13}$ &  {119740} & {11137472} &  {11041759} & {113491483} & {22014(14)} & {1291.3} \\ 
\hline 
\hline 
Total & 541948 &  54323582 & 48647958  & 518756035  &  88073  & {6141.9} \\
\hline
\end{longtable}
The total CPU wall time is then {1 day 27 minutes and 53 seconds}, and the total memory usage is {6.142} Gigabytes. The maximal absolute error on eigenvalues {does not} go over 0.47 $\rm{cm}^{-1}$ giving a maximal relative error lower than $4*10^{-5}$.
As a matter of comparison the reference calculation were done with a total memory usage of 128 gigabytes and 3 days time on a 24 cores computer meaning that the CPU wall time is much larger. Less accurate results (4-5 $\mathrm{cm}^{-1}$ error on higher frequencies) are achieved with HRRBPM \cite{Thomas2016} with a memory usage of 14.6 gigabytes and a CPU wall time of 8.7 days.
\subsection{$\rm{C_3H_3NO}:$ Oxazole}
 The PES was constructed using the adaptive density-guided approach (ADGA) introduced by Sparta et al  \cite{Sparta2010,Sparta2009,Sparta2009_2}.
The force constants, equilibrium geometry and normal coordinates where extracted from Madsen et al \cite{Madsen2017}.
In their work they describe the construction of oxazole PES at CCSD(T)/cc-pVTZ level for the one-mode part and MP2/cc-pVTZ for the two-mode part. The three-mode part is extrapolated from the two-mode surface using MP2/cc-pVTZ gradients.
The number of terms is 146 for the one mode, 4786 for the two modes and 4335 for the three modes couplings.
\paragraph{Harmonic frequencies}~\newline
\begin{eqnarray*}
 \nu_{1} : 603.8 \rm{cm}^{-1}, \  \nu_{2} : 644.2 \rm{cm}^{-1}, \  \nu_{3} : 748.9 \rm{cm}^{-1}, \\
 \nu_{4} : 832.2 \rm{cm}^{-1}, \ \nu_{5} : 849.7 \rm{cm}^{-1}, \ \nu_{6} : 901.8 \rm{cm}^{-1}, \\
 \nu_{7} : 913.8 \rm{cm}^{-1}, \ \nu_{8} : 1071.8 \rm{cm}^{-1}, \ \nu_{9} : 1109.2 \rm{cm}^{-1}, \\
 \nu_{10} : 1106.9 \rm{cm}^{-1}, \ \nu_{11} : 1181.2 \rm{cm}^{-1}, \ \nu_{12} : 1263.5 \rm{cm}^{-1}, \\
 \nu_{13} : 1348.3 \rm{cm}^{-1}, \ \nu_{14} : 1533.4 \rm{cm}^{-1}, \  \nu_{15} : 1570.4 \rm{cm}^{-1}, \\
 \nu_{16} : 3275.6 \rm{cm}^{-1}, \ \nu_{17} : 3286.0 \rm{cm}^{-1}, \ \nu_{18} : 3309.5 \rm{cm}^{-1}. \\
\end{eqnarray*}
{The maximal harmonic frequency Freq0Max=20000 $\mathrm{cm}^{-1}$ associated with MaxQLevel=10, gives the rounded pruning condition
\[\left\{\begin{array}{l}\displaystyle \vect{b} \in \mathbb{N}^{18}, \ \vect{b}\leq (10, 10, 10, 10, 10, 10, 10, 10, 10, 10, 10, 10, 10, 10, 10, 6, 6, 6)\\
\displaystyle  b_1 + b_2 + b_3 + b_4 + b_5 + b_6 + 2b_7 + 2b_8 + 2b_9 + 2b_{10} + 2b_{11} + 2b_{12} \\
 + 2b_{13} + 3b_{14} + 3b_{15} + 5b_{16} + 5b_{17} + 5b_{18} \leq 33 \end{array} \right\}.\]
}~\newline
\paragraph{Results}~\newline
\begin{longtable}{|c|c|c|c|c|c|}
\caption[]{Oxazole anharmonic frequencies for fundamental targets separated by an horizontal bar.}\\
\hline
Eigenvalue & Frequency & Relative & {$\Delta E$ \eqref{eqn:ReziduD}} & Assignment & Experimental  \\
 number &   & Residue & &(component) & values \cite{Pouchan1974} \\ 
\hline 
   0\footnote{{Watson term $-\frac{1}{8}\sum _{\alpha =1}^{3}\mu _{\alpha \alpha }= -0.2052$}} &   12559.9032   &  0.0046 &-0.2770 & $\nu_{0}(0.95)$ & 12457.5 \\ 
   1   &   592.8494   &  0.0074 &-0.9701&  $\nu_{1}(0.95)$ & 607 $(A^{''})$ \\  
   2   &   631.2242   &  0.0075 &-0.9924&  $\nu_{2}(0.95)$  & 647 $(A^{''})$  \\   
   3   &   727.3345   &  0.0072 &-0.8395&  $\nu_{3}(0.93)$  & 750 $(A^{''})$  \\   
   4   &   795.0331   &  0.0077 &-0.9749&  $\nu_{4}(0.90)$  & 830$(A^{''})$   \\   
   5   &   827.7750   &  0.0070 &-0.8230&  $\nu_{5}(0.90)$  & 854  $(A^{''})$ \\  
   6   &   884.4607   &  0.0074 &-0.9951&  $\nu_{6}(0.92)$  & 899 $(A^{''})$ \\   
   7   &   894.7569   &  0.0071 &-0.9031&  $\nu_{7}(0.91)$  & 907 $(A^{''})$  \\   
   8   &   1031.4634   &  0.0064 &-0.7109&  $\nu_{8}(0.90)$  & 1046 $(A^{'})$  \\   
   9   &   1063.2165   &  0.0067 &-0.7752&  $\nu_{10}(0.82)$  & 1078 $(A^{'})$ \\   
   10   &   1075.3896   &  0.0069 &-0.8504&  $\nu_{9}(0.82)$  & 1086 $(A^{'})$ \\   
   11   &   1123.9852   &  0.0066 &-0.7687&  $\nu_{11}(0.89)$  & 1139 $(A^{'})$ \\   
   13   &   1217.7686   &  0.0066 &-0.7716&  $\nu_{12}(0.94)$  & 1252 $(A^{'})$ \\   
   16   &   1302.4880   &  0.0078 &-1.1508&  $\nu_{13}(0.90)$  & 1324 $(A^{'})$  \\   
   24   &   1481.3806   &  0.0079 &-1.2277&  $\nu_{14}(0.86)$  & 1504 $(A^{'})$ \\   
   27   &   1521.2312   &  0.0074 &-1.0519&  $\nu_{15}(0.91)$  & 1537 $(A^{'})$ \\  
\hline   
\hline
   585   &   3125.9955   &  0.0086 &  -2.3563 &  $\nu_{16}(0.76)$ & 3141 $(A^{'})$\\
   608   &   3146.9603   &  0.0098 & -2.6857  &  $\nu_{17}(0.82)$ & 3144 $(A^{'})$\\
   618   &   3159.5778   &  0.0087 &-2.1459 &  $\nu_{18}(0.81)$ & 3170 $(A^{'})$\\
\hline
\end{longtable}
\paragraph{Performances summary}~\newline 
\begin{longtable}{|c|c|c|c|c|c|c|}
\caption[]{Performances summary on Oxazole molecule. Each screened states are separated by an horizontal bar. The CPU wall time is in second with the total number of iterations indicated in parenthesis.
In both cases, NAdd=200, EtaComp=3, Freq0Max=20000, {ThrKX=1. EpsRez=(0.008,0.01) respectively for each group.}}\\
\hline
Target(s) & Final & Final  &   Final  &  Final  &CPU Wall time(s)& Memory \\
 &  size of $B$ & size of $B_{\smalltt{S}}$ & $nnz(H_{\smalltt{B}})$ &  $nnz(H_{\smalltt{SB}})$ & (Iterations) & usage (MO)\\ 
\hline
$\nu_0-\nu_{15}$ & 145820 & 27468841 & 87344774 &  299972877 & 74665(10)& {5424.3}\\
\hline
$\nu_{16}-\nu_{18}$ & 143916 & 29724836 & 60788880 &  285617698  & 170658(19)& {4933}\\
\hline
Total & 289736 & 57193677 & 148133654 & 585590575 & 245323 & {10357.3} \\
\hline
\end{longtable}
The total cpu wall time is 2 days 20 hours 8 minutes 43 seconds. A significantly {higher latency} for the second group of targets $\nu_{16}-\nu_{18}$ principally comes from the additional number of iterations.
To a lesser extent, there is also the constraint to calculate the eigenvalues starting from the extremities of the spectrum as in Lanczos algorithm. A {Jacobi-Davidson eigensolver \cite{Sleijpen2000,G.Sleijpen1996,Petrenko2017a}} or polynomial filtering techniques \cite{Sorensen1997,Karlsson2007,Fang2012a,Zhou2007} could be more adapted.

\begin{figure}[!htbp]
\center
\includegraphics[width=0.9\textwidth]{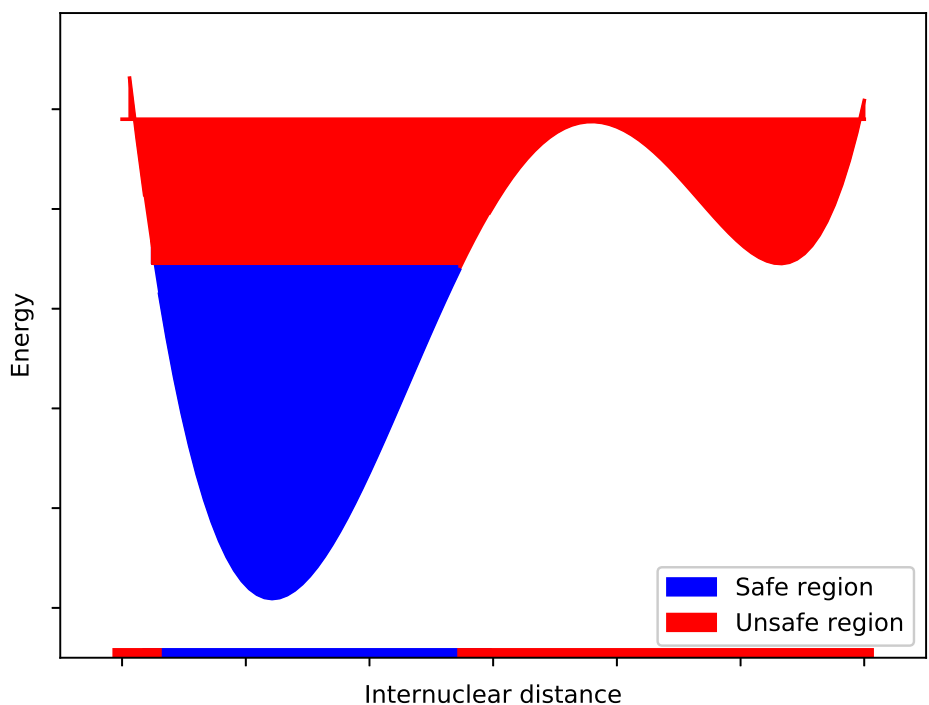}
\caption{{Fictitious potential energy surface depending on one internuclear coordinate showing 2 regions. The blue one is quasi quadratic when viewed from the internuclear distance and corresponding energy. In the red area, additional interactions are brought by the second critical point provoking an uncontrollable perturbative effect on the variational solutions.}} 
\label{Fig:PollutedPES}
\end{figure}

{The energy barrier Freq0Max has voluntary been lowered down, due to important successive shifting or oscillation of the position of the tracked eigenvalues. This phenomenon usually occurs when the PES is no longer locally quadratic for some particular configurations.
It is well illustrated in the case of double well potentials \cite{Wilbur1994a,Lourderaj2008} and in figure \ref{Fig:PollutedPES} showing a fictitious PES oscillating beyond a given spatial region. Another way to get around this exception would be to use localized basis functions such as distributed Gaussians \cite{Hamilton1986} directly enabling a restriction of the spacial area.
}
\pagebreak
\section{Input parameters}
\paragraph{Presentation}
The key words are case insensitive and should start at the beginning of each line of the input file. 
No specific order of apparition is required. Comments are indicated with the symbols '/' or '@'.
The potential energy file should also be present in the directory where is executed DVCI (cf PESType\ref{PESType}).
A minimal input file looks like~:
\begin{verbatim}
NMode 6                  /  Number of normal coordinates
PESType 1                /  PES type of coefficients
OutName N2H2             /  Extension name for output files.
PESName N2H2_PES.in      /  Name of the file of the PES.
Memory 80               /  Maximal allocated memory in mega octets.
\end{verbatim}
\paragraph{Detailed list of the key words}
\begin{enumerate}[label=$\diamond$\arabic*]
\item{NMode} ~\newline Designs the number of mass weighted normal coordinates.~\newline 
 ${\rm NMode}=3*N_A-6$. Where $N_A$ stands for the number of atoms of the molecule.
\item{DoRot}\label{DoRot} ~\newline
Indicate if {Coriolis} corrections should be added to the Hamiltonian.~\newline
In the PES file, the section to be filled should start and end with words ~\newline COORDINATES ENDCOOR and each different field must be separated by an exclamation mark.
\begin{enumerate}[label=\textbf{\arabic*}]
\item[$\bullet$]{If $= 0$}~\newline then $\displaystyle\mathcal{H}=\mathcal{H}_{vib}(\vect{q})$\eqref{eqn:HVib}.
If no section COORDINATES is written in the PES file then DoRot is automatically set to 0.
\item[$\bullet$]{If $> 0$ and even}\label{NCoor}~\newline then $\displaystyle\mathcal{H}=\mathcal{H}_{vib}(\vect{q})+\mathcal{H}_{{CC}}(\vect{q})$ \eqref{eqn:HRot}. 
The equilibrium geometry in Bohr, atomic masses in electron rest mass, and normal coordinate eigenvectors should be indicated in the section COORDINATES like in the following example where the values has been extracted from Madsen \& al\cite{Madsen2017}.
\begin{verbnobox}[\small]
COORDINATES
! Equilibrium geometry (bohr)
+1.188227703663817e+00  +1.871902685046820e+00  +5.700041831365050e-16 /C
-1.366889823622856e+00  +1.691283181574289e+00  +5.673619333132633e-16 /C
-1.998417882238274e+00  -8.244795338089052e-01  -2.155218936400323e-16 /O
+2.684304163103964e-01  -2.022882150581233e+00  -6.432553562388595e-16 /C
+2.229151382700317e+00  -5.583543727400242e-01  -2.215072662048790e-16 /N
+2.352687276664275e+00  +3.539113911722189e+00  +8.792950593981537e-16 /H
+2.239397718366979e-01  -4.057044206724038e+00  -1.063440276037351e-15 /H
-2.901750851724292e+00  +3.020857567289711e+00  +7.990414399553623e-16 /H
! Masses(me)
21874.6618172 /C
21874.6618172 /C
29156.9456749 /O
21874.6618172 /C
25526.0423547 /N
1837.15264562 /H
1837.15264562 /H
1837.15264562 /H
! Mode0:    X                     Y                     Z
 +3.231683769903235e-17 +6.115580033151089e-17 +4.025196381179206e-01
 +2.743657157380967e-17 +4.590882496215692e-17 -5.186966592183442e-01
 +1.991732529600559e-17 +5.710133055001405e-17 +5.169293296045854e-01
 -7.257396408556236e-17 +7.324385585074392e-17 -1.938605777276555e-01
 -9.053243042419710e-17 +2.384482767132475e-18 -1.671562436764069e-01
 +3.416768858945461e-17 -1.546457080242301e-17 +2.188043144086224e-01
 -1.950243964435382e-17 -1.457271158617074e-16 -2.002715238252209e-01
 -5.645471816646615e-17 -3.099000577024135e-17 -3.849787518532105e-01 
! Mode1      
 -4.471096934999722e-17 -2.133222140486714e-16 +3.245666277578175e-01
 -7.195147510789707e-18 -1.776354539431898e-16 -3.130286914609319e-02
 -1.161675776799252e-16 -2.465033916085211e-16 -2.713644979495499e-01
 -7.304974442800625e-17 +1.130401139043093e-16 +4.555118862530116e-01
 +1.980788418374012e-17 -1.509309233138445e-16 -5.878836462126088e-01
 +3.871182981501824e-17 +3.894914167531566e-17 +2.788008333658181e-01
 -1.775227773837671e-18 +1.687750614789954e-16 +4.342956786268967e-01
 -9.998627049911543e-18 -1.727718015508009e-16 -2.443444074954252e-02 
! Mode2      
 -1.362124303938439e-16 +1.648858093263572e-17 +7.985191937307874e-02
 +9.223533160690561e-17 -1.520416449194739e-16 +4.427598081406853e-01
 -3.106447728140722e-16 -2.024361724665231e-16 -7.334352062590022e-02
 +3.248565849336364e-17 +2.181702164863730e-17 -1.148459584461199e-01
 +3.231452029113849e-17 +1.995473253650141e-16 -3.854511885546462e-02
 -1.704399688795554e-16 +6.014500621640033e-17 -3.252722203195085e-01
 +1.114007103292430e-16 +8.683057684777304e-18 +1.582562080820018e-01
 +3.567043855971248e-16 +2.236379927154288e-16 -8.041677585822095e-01 
! Mode3      
 -8.530348729807672e-17 +5.064459244391696e-16 +2.099386275547177e-01
 -2.860512641462178e-17 +2.849372440086422e-16 -1.949978814638483e-02
 +9.671777362477948e-16 +3.228924378351529e-16 -1.437578533863218e-01
 -3.535112014919970e-16 -6.871677835537049e-16 +4.816962590153900e-01
 -6.328554966609436e-16 +1.463685147682383e-16 -1.649351614432971e-01
 +3.642789510512844e-16 +2.417462616289285e-18 -4.324245782423197e-01
 -5.058760423068888e-17 -3.092952891160942e-16 -6.990373338704934e-01
 -6.384663222086937e-16 -5.923791764375996e-16 -3.242248209522854e-04 
       .                       .                       .
       .                       .                       .
       .                       .                       .
 ENDCOOR
\end{verbnobox}
In accordance with the Wilson method\cite{Wilson1955}, the normal coordinates are built from a set of $3N_A-6$ eigenvectors $\displaystyle (Q_{i_{\alpha}})_{\ i \in \{1,\ldots ,N_A\}, \ \alpha\in \{1,2,3\}}$ of the Hessian matrix \begin{equation}\label{eqn:Hessian}\left(\frac{1}{\sqrt{m_i} \sqrt{m_j}}\frac{\partial^2\mathcal{U_K}}{\partial x_{i_{\alpha}} \partial x_{j_{\beta}}}\right)_{ (i,j) \in \{1,\ldots ,N_A\}^2, \ (\alpha,\beta)\in \{1,2,3\}^2}\end{equation} derived from mass weighted displacements ($m_i$ mass of nucleus $i$)
\[\Delta y_{i\alpha}=\sqrt{m_i}\left(x_{i_\alpha}-Xeq_{i_\alpha}\right), \ \ i \in \{1,\ldots ,N_A\}, \ \alpha\in \{1,2,3\}, \] at the equilibrium geometry $\vect{Xeq}\in \reel^{3N_A}$. The corresponding eigenvalues are the harmonic frequencies. 
\item[$\bullet$]{If $>0$ and odd}\label{Rot0}~\newline
then $\displaystyle\mathcal{H}=\mathcal{H}_{vib}(\vect{q})+\mathcal{H}_{{CC}}(\vect{q})$ and the non mass weighted normal coordinate eigenvectors \[\displaystyle (Q_{i_{\alpha}}/\sqrt{m_i})_{\ i \in \{1,\ldots ,N_A\}, \ \alpha\in \{1,2,3\}}\] should be written instead of the classical ones. 
\end{enumerate}
\item{PESType}\label{PESType} ~\newline
Format of data's for the {multivariate} PES.
\begin{enumerate}[label=\textbf{\arabic*}]
\item[$\bullet$]{If PESType $= 0$} then the force constants $K_{\vect{c}}$ are expressed for dimensionless normal coordinates $(q_n=Q_{n}/\sqrt{\nu_n})$ and supplied in $\rm cm^{-1}$.
Regarding the format of the PES it starts and ends with the key words~\newline FORCEFIELD ENDFF. For NM normal coordinates, NM integers should be shown before the actual value of the force constant:
\begin{verbatim}
FORCEFIELD
 2 0 0 0 0 0 , 664.213550943134237
 4 0 0 0 0 0 , 4.335282791860437
 6 0 0 0 0 0 , -0.471897107116644
 0 2 0 0 0 0 , 675.140549094012272
 0 4 0 0 0 0 , 7.072599090662686
 2 0 1 0 1 0 , -12.115726349955748
 2 0 1 0 2 0 , 1.215724777835937
 2 0 1 0 3 0 , 0.161839021650279
 2 0 1 0 1 2 , 0.800398391535361
 2 0 1 0 0 2 , 0.662431742378760
 2 0 0 1 0 0 , 11.492641524810505
 2 0 0 2 0 0 , 0.117652856544075
 2 0 0 3 0 0 , -0.819356186630299
 . . . . . . ,      .
 . . . . . .        .
 . . . . . .        .
ENDFF 
\end{verbatim}
Here it means that the first term is the one in front of $q_1^2$ namely $\nu_1/2$,~\newline and for the last showed line 2 0 0 3 0 0, we are dealing with the force constant $K_{2, 0, 0, 3, 0, 0}=-0.819356186630299 \ \rm{cm}^{-1}$ in agreement with the monomial $q_1^{2}q_4^{3}$. 
\item[$\bullet$]{If PESType $\geq 1$} then the derivatives are provided in atomic units, and for a Taylor expansion around the equilibrium position we have the correspondence
\begin{equation}\label{eqn:Taylor}
K_{c_1\ldots c_{\mathrm{NM}}}=\frac{1}{c_1!c_2!\ldots c_{\mathrm{NM}}! \prod_{n=1}^{\mathrm{NM}}\sqrt{\nu_{n}}^{c_n}}
\displaystyle\frac{\partial^{c_1,\ldots , c_{\mathrm{NM}}} \mathcal{U_K}}{\partial Q_{1}^{c_1} \ldots \partial 
Q_{\mathrm{NM}}^{c_{\mathrm{NM}}} }*\mathrm{HartreeToCM},
\end{equation}
where HartreeToCM is a converting factor from Hartree to $\rm{cm}^{-1}$ and $\nu_n=\sqrt{\frac{\partial^{2} \mathcal{U_K}}{\partial Q_{n}^{2} }}$.
The format of PES file is the same as the one supplied by the PyPES\cite{PyExtens} library namely:
\begin{verbatim}
          FORCEFIELD
          [0,0,0,0,3,4 , 1.86861408859e-11],
          [0,0,0,0,4 , -3.47804520495e-09],
          [0,0,0,0,4,4 , 3.2867316136e-10],
          [0,0,0,0,5,5 , 3.53998391655e-10],
          [0,0,1,1 , 6.40569999931e-09],
          [0,0,1,1,1,1 , -3.21892597605e-11],
          [0,0,1,1,1,5 , -4.73206802979e-12],
          [0,0,1,1,2 , 1.16300104731e-11],
          [0,0,1,1,2,2 , -5.81559507099e-12],
          [0,0,1,1,2,3 , 9.06462603669e-12],
          [0,0,1,1,2,4 , -2.30405700341e-12],
          [0,0,1,1,3 , -1.89925822496e-10],
          [0,0,1,1,3,3 , -1.6879336773e-11],
          [0,0,1,1,3,4 , 1.08730049094e-11],
           . . . . . . ,      .
           . . . . . .        .
           . . . . . .        .
           ENDFF 
\end{verbatim}
The repetitions are to be associated with a derivative order when coordinates are numbered starting from zero. For example the first line means
\[\frac{\partial^{4}}{\partial Q_1^{4}} \frac{\partial}{\partial Q_4}  \frac{\partial}{\partial Q_5} \mathcal{U_K}=\mathrm{1.86861408859 \ * \ 10^{-11} \ \ a.u.} \]
\end{enumerate}
\item{{PESName}}\label{PESName}\newline{Name of the file that contains the force constants or derivatives.}
\item{ThrPES}\label{ThrPES}~\newline
Threshold for PES force constants or derivatives. Default value is {the double precision error machine $\simeq 2*10^{-16}$}.
\item{EpsRez}\label{EpsRez}~\newline For eigenvectors of $H_{\smalltt{B}}$ \eqref{eqn:HRRJ} $\vect{X}_{\ell}, \ \ell\in \{\mathrm{Targ}\}$, it is the maximal accepted relative residue \[\max_{\ell\in \{\mathrm{Targ}\}}\frac{\Vert H_{\smalltt{SB}}\vect{X}_{\ell}\Vert}{E_{\ell}}\] before the algorithm stop. They are built from the MVPs \eqref{eqn:Complet}. The default value is $6*10^{-3}$.
\item{ThrMat}\label{ThrMat}~\newline
Minimal allowed absolute value of coefficients of $H_{\smalltt{B}}$. Default is {the double precision error machine $\simeq 2*10^{-16}$}.
The matrix coefficients are computed with the full operator $\mathcal{H}=\mathcal{H}_{vib}+\mathcal{H}_{{CC}}$. If DoRot=0, only $\mathcal{H}_{vib}$ will be considered.
\item{MaxQLevel}\label{MaxQLevel}
  ~\newline This is the common maximal quantum level for the whole space $B\oplus B_{\smalltt{S}}$. 
It increases together with distances between nucleus in motion and then should carefully be chosen conforming to the spacial region where the potential energy is still correctly represented and has no more than one {critical point}. Each {upper} level {$d_n$} on normal coordinate $n$ will be adjusted with Freq0Max\ref{Freq0Max} as followed:{\[d_n=\min\left(\lfloor \frac{\mathrm{Freq0Max}}{\nu_n}\rfloor,\mathrm{MaxQLevel}\right)\]}
\item{Freq0Max}\label{Freq0Max}~\newline Maximal allowed harmonic {frequency}\footnote{{$\displaystyle=\sum_{n=1}^{\mathrm{NM}} b_n*\nu_n$}} in $B\oplus B_{\smalltt{S}}$. Default value is 30000.
\item{ThrKX}\label{ThrKX}~\newline
In operator $\mathcal{H}^*$\eqref{eqn:DualH} only the increments $\vect{e}\in \mathrm{LFF}^*$ verifying \footnote{$Z_{ijkl}$ is defined equation \eqref{eqn:HRot}}
\[\sum_{\vect{c}\in \{\mathrm{LFK}(\vect{e})\}} |K_{\vect{c}}|+\sum_{(i,j,k,l)\in \{\mathrm{LCI}(\vect{e})\}} |Z_{ijkl}|>\mathrm{ThrKX},\]  
will be acceptable to generate the secondary space $B_{\smalltt{S}}$ and residual vectors.
It should be strictly positive. Default value is 1. 
\item{NAdd}\label{NAdd}~\newline
It is the minimal number of basis functions per non converged target states to be added for next iteration. They are chosen from maximal components (in absolute value) of the residual vectors \eqref{eqn:Complet} \begin{equation}\label{eqn:SetRez}\left\{(H_{\smalltt{SB}}\vect{X}_{\ell})_{\vect s}, \ \ell \in {\mathrm{NotConv}}, \ \vect{s}\in B_{\smalltt{S}}\right\},\end{equation} 
where NotConv designs the set of non converged eigenpairs
\begin{equation}\label{eqn:NC}\mathrm{NotConv}=\{\ell\in \{\mathrm{Targ}\}, \ \frac{||H_{\smalltt{SB}}\vect{X}_{\ell}||}{|E_{\ell}|} >\mathrm{EpsRez}\}.\end{equation} 
To accelerate convergence, NAdd is multiplied by $\mathtt{i}+1$, where $\mathtt{i}$ designates the iteration number.
\item{EtaComp}\label{EtaComp}~\newline
The new added basis functions are selected from all the components of the residual vectors \eqref{eqn:Complet} greater than \[\frac{1}{\mathrm{EtaComp}*\mathrm{NNotConv}}\sum_{\ell \in \mathrm{NotConv}}\Vert H_{\smalltt{SB}}\vect{X}_{\ell}\Vert_{\infty}\] where NotConv \eqref{eqn:NC} and NNotConv respectively stand for the set of non converged tracked eigenpairs of $H_{\smalltt{B}}$ and its cardinal. EtaComp should be greater than one, this {turns to be a guaranty} that at list one component per non converged residual vector will be picked up.
\item{MaxAdd}\label{MaxAdd}~\newline Limit for number of basis functions to add at each iteration. Default value is 1000.
\item{TargetState}\label{Target}~\newline
 It indicates the maximal component of the eigenvectors of $H_{\smalltt{B}}$ that should be assigned to the targets matching with a multi-index array (cf figure \ref{Fig:KInd}).
Except for $\vect{0}_{\rm NM}$ symbolized by 0(1), only its non zeros should be indicated with the characters $d(n)$ separated by a comma, where $d$ stands for the degree of the Hermite function and $n$ the normal coordinate.
Alternatively TargetState can be followed by the label 'Fund' if the targets are the fundamentals and the ground state i.e $1(n), \ n=\{1, \ldots , \mathrm{NM}\}$ and $0(1)$.
If $0(1)$ is not part of the targets then the zero point energy should be provided in $\rm cm^{-1}$ via the parameter GroundState \ref{GState}.
\item{ThrCoor}\label{ThrCoor}~\newline
Any eigenvector coordinate of  $H_{\smalltt{B}}$ bigger (in absolute value) than this threshold and assigned to one of the targets, will be integrated into the iterative process and have its residual vector \eqref{eqn:Complet} calculated. 
 In output, will be showed only the assignments of components larger than ThrCoor. 
\item{AddTarget}\label{AddTarget}~\newline
Sometimes, different eigenvectors point to the same maximal components. Then the actual number of targets is bigger than the one specified by the user. So it allocates additional arrays to correct this increasing. The default value is 2.
\item{GroundState}\label{GState}~\newline
Zero point energy required when it is not calculated (i.e not part of the targets).
It can also be adopted as reference to printout the anharmonic frequencies.
\item{MinFreq, MaxFreq}\label{MinFreq}~\newline
Frequencies in $\mathrm{cm}^{-1}$ specified to make converge all the eigenvalues within the interval \[\rm [MinFreq+GroundState,MaxFreq+GroundState]\] when no target is indicated. If MinFreq is greater than zero, the value of GroundState should be supplied, else it will be computed. If MaxFreq is not given in input, then it will be set to the maximal harmonic frequency of tracked states for the initial subspace construction and to Freq0Max\ref{Freq0Max} afterwards. Default values are [-100,4000].
\item{Kappa}\label{kappa} ~\newline
{Empirical elongation factor accounting the maximal gap between an harmonic and a converged energy number $\ell$ when ordered like in \eqref{eqn:num}}. Its default value is 1.2 but it is automatically augmented to 1.3 when maximal target frequency is greater than $3000 \ \rm{cm}^{-1}$.
\item{Memory}\label{Memory}~\newline
Total allocated memory in megabytes for the slots occupied by the eigensolver, the matrices $(H_{\smalltt{B}}, \ H_{\smalltt{SB}})$, the multi indexes $(B, \ B_{\smalltt{S}})$, the PES, the local force fields and corresponding positive increments $\left\{\mathrm{LFF}(\vect{e}), \ \vect{e}\in \mathrm{LFF}^*\right\}$\eqref{eqn:LFF}. This value will be used to set up the upper limit of basis functions SizeActMax that is appraised taking into account the array shrinkage factors KNREZ\ref{KNREZ}, KNNZ\ref{KNNZ} and KNZREZ\ref{KNZREZ}. 
\item{KNNZ}\label{KNNZ}~\newline Sparsity factor for $H_{\smalltt{B}}$.
The maximal number of non zero coefficients in $H_{\smalltt{B}}$ will be \[\rm NNZActMax=KNNZ*SizeActMax*NXDualHTrunc.\] Where NXDualHTrunc is the upper limit of excitations in $\mathcal{H}^*$ after truncation with ThrKX\ref{ThrKX}. Default value is 0.03. Should be in ]0,1]. 
\item{KNREZ}\label{KNREZ}~\newline Multiplicative factor of the maximal size of the residual space \[\rm SizeRezMax=KNREZ*SizeActMax*(NXDualHTruncPos-1).\] Where NXDualHTruncPos-1 is the number of raising excitations in operator $\mathcal{H}^*$ after truncation with ThrKX\ref{ThrKX} (the first excitation being zero). Default value is 0.2. Should be in ]0,1].
\item{KNZREZ}\label{KNZREZ}~\newline Shrinking factor for the maximal number of pointers on the non zeros of $H_{\smalltt{SB}}$ \[\rm NNZRezMax=KNZREZ*SizeActMax*NXDualHTrunc.\] Where NXDualHTrunc is the upper limit of excitations in operator $\mathcal{H}^*$ after truncation with ThrKX\ref{ThrKX}. 
KNZREZ will be settled to zero when DoGraph=0.
\item{DoGraph}\label{DoGraph}
\begin{itemize}\item[$\bullet$If$=0$] The MVPs $H_{\smalltt{SB}}\vect{X}_{\ell}, \ \ell\in \{\mathrm{Targ}\}$ are fully calculated by browsing $\mathcal{H}^*(B)$ instead of $\mathcal{H}^*(A)$ in \eqref{eqn:MVPPart}.
\item[$\bullet$If$>0$] The row indexes and column pointers of the coupled elements of $H_{\smalltt{SB}}$ 
are stored in CSC\footnote{Compressed Sparse Column} format to complete $\widetilde{H_{\smalltt{SB}}\vect{X}_{\ell}}, \ \ell\in \{\mathrm{Targ}\}$ in \eqref{eqn:MVPPart} for the missing entries \eqref{eqn:Complet}. This option necessary increases memory requirement. The expense is about 40\% greater in memory and 40\% smaller in CPU time compared with DoGraph=0. 
\end{itemize}
Default value is 1.
\item{MaxEV}\label{MaxEv}~\newline 
Maximum eigenvalues to be computed when counted from the smallest one.  
This number is adjusted to the size of the initial subspace minus one when it is actually larger than the latter. The eigensolver uses the Mode 1 and option WHICH='LM' of ARPACK subroutine \href{http://www.caam.rice.edu/software/ARPACK/UG/node136.html}{DSAUPD}. The greatest magnitude eigenvalues are computed on the shifted matrix \[H_{\smalltt{B}}^{'} =H_{\smalltt{B}}-\mathrm{Shift}*I_{\smalltt{B}},  \ \mathrm{Shift}=\sum_{n=1}^{\rm NM}\left[\frac{1}{2}+\mathrm{MaxQLevel}(n)*\nu_n\right],\]
where $I_{\smalltt{B}}$ designates the identity matrix.
Default value is 30.
\item{DeltaNev}\label{DeltaNev}~\newline 
Reduce the number of wanted eigenvalues as
 \[ \rm  MaxEV=Min(MaxEV, MaxScreen+DeltaNev),\] where MaxScreen is the higher position of the targets that tends to decrease with iterations. The purpose is to lighten the computational effort on the eigensolver. Default value is 1000.
\item{MAXNCV}\label{MAXNCV}~\newline
This is the maximal number of Lanczos basis vectors generated at each iteration in \href{http://www.caam.rice.edu/software/ARPACK/UG/node136.html}{DSAUPD} subroutine. Default value is 2*MaxEv\ref{MaxEv}.
\item{Tol}\label{Tol}~\newline
 Stopping criterion for the relative accuracy of the Ritz values in \href{http://www.caam.rice.edu/software/ARPACK/UG/node136.html}{DSAUPD} subroutine. 
Default value is $10^{-8}$.
\item{RefName}\label{RefName}~\newline Name of the input text file holding a floating point number at the beginning of each line to compare with the final results. The printed error is the difference between one of this value and the closest calculated frequency. Then it should manually be corrected when this correspondence is not true.
\item{Verbose}\label{Verbose}~\newline
When non equal to zero, it allows to print additional informations such as intermediate CPU times, position of targets in initial space, the center of mass, the moment of inertia and the characteristics of the dual operator.
\item{OutName}\label{OutName}~\newline 
Extension for output file names created when PrintOut$\neq 0$ (cf\ref{PrintOut}).
\item{PrintOut}\label{PrintOut}~\newline 
\begin{enumerate}
\item[$\bullet$]{If {$= 1$}}: All the final basis set {and the components of the eigenvectors will be saved in the files OutName-FinalBasis.bin and OutName-Vectors.bin}.
 These informations could be employed to compute infrared  intensities in the final basis set {with the module Transitions that evaluates the quantities
\begin{equation}\label{eqn:Tr}\braket{\Psi_0|\mathcal{O}|\Psi_\ell}, \ \ell\in \{\mathrm{Targ}\}, \end{equation}
where $\mathcal{O}$ is a given operator that should have the same format than the PES used for DVCI. $(\Psi_0,\Psi_\ell)$ are the wave functions of the ground and target state $\ell$ respectively.
Under the transition moment\eqref{eqn:Tr} is also printed the difference of corresponding eigenvalues
\[F_\ell=E_\ell-E_0,\]
permitting to retrieve the infrared intensity when the dipole moment vector $\boldsymbol{\mu}(\vect{q})$ is supplied as a function of the normal coordinates through the formula \cite{Seidler2007}
\[I_\ell=\frac{N_A}{6C^2\epsilon_0\hbar^2} F_\ell |\braket{\Psi_0|\boldsymbol{\mu}(\vect{q})|\Psi_\ell}|^2 (m_0 - m_\ell ).\]
$C$ is the speed of light, $\epsilon_0$ the vacuum permittivity, $\hbar$ the reduced Planck constant, and $(m_0 - m_\ell)$ the difference of Mole fractions that is usually set up to one at zero temperature.
The parameters of the input file are the same as DVCI and PESName\ref{PESName} should be replaced by the name of the file containing operator $\mathcal{O}$.}
  \item[$\bullet$]{If {$=2$}}: {The last iteration can be replayed by using exactly the same input file as DVCI with the executable called FinalVCI.}
  \item[$\bullet$]{If {$>2$}}: {The size of the reference space defined with the pruning condition\eqref{eqn:Prun} and maximal quantum levels \eqref{eqn:MaxQ} can be evaluated with the program Transitions.}  
\item[$\bullet$]{If $= 0$}: No additional output file is created.
\end{enumerate}
Default value is 0.
{\item{EvalDeltaE}\label{DeltaE}\newline If $\neq 0$ the correction energies $\Delta E$ \eqref{eqn:ReziduD} will be evaluated and printed at the end. Default value is 0.}

\end{enumerate}

\section{Conclusion}\label{Conclu}
In this work has been presented a new algorithm to track specific states of molecular spectrum approaching the variational limit with a minimal usage of memory. Harmonic oscillator properties together with second quantization formulation were adopted to build a novel assemblage of structures available for dynamic subspace enrichment. 
The resulting code has shown challenging performances and could obviously be applied for bigger systems that the ones studied in here.
Remains the possibility to adapt the method for different implementations of internal coordinates already available in a software like TROVE \cite{TROVE}. 
 {The overall construction might also be extended to other kind of basis functions if analytical calculation rules can be factorised for a given form of potential energy that should minimally be written as a sum of product.} 
\section*{Acknowledgments}
I would like to thank professor T.Carrington for providing me the potential energy surface of ethylene oxide.
\section*{Appendix : Hermite function analytical formulas}\label{App}
In one dimension, Hermite functions verify
\begin{equation}\label{eqn:Herm}\int_{\reel} \psi_{b}(q) q^{d_1} \frac{\partial^{d_2}}{\partial q^{d_2}} \psi_{b+e}(q) dq  \neq 0 \ \mathrm{if \ } \exists \ t\in \mathbb{N}, \ |e|=d_1+d_2-2t,\end{equation}
as well as for the switched product $\frac{\partial^{d_1}}{\partial q^{d_1}}q^{d_2}$. ~\newline
It is easily demonstrable with recurrence relations \cite{gradshteyn2007,GaborSzego1967}
  \begin{multline}\label{eqn:Recur}\psi _{b}'(q)={\sqrt {\frac {b}{2}}}\psi _{b-1}(q)-{\sqrt {\frac {b+1}{2}}}\psi _{b+1}(q)~,\\
 q\;\psi _{b}(q)={\sqrt {\frac {b}{2}}}\psi_{b-1}(q)+{\sqrt {\frac {b+1}{2}}}\psi _{b+1}(q),\\ \end{multline}
and can directly be related to the definition of operators \eqref{eqn:Op2}.
For the coefficients \[\braket{[q^d]}_{b,s}= \braket{\psi_{b}  (q)|q^d|\psi_{s}  (q)},  \ (b,s) \in \{0, \ldots, \mathrm{Dim}\}^2, \ d\geq 1,\] the following property applies
\[ \braket{[q]^d}_{b,s} = \braket{[q^d]}_{b,s} \ , \ \forall  (b,s) \in \{0,\ldots, \mathrm{Dim}-d+1\}^2. \]
where
\[[q]=\begin{bmatrix} 0 & \sqrt{\frac{1}{2}} & 0      & \cdots & \cdots & 0 \\ \sqrt{\frac{1}{2}} & 0 & \sqrt{\frac{2}{2}} & \ddots & \ddots & \vdots \\  0     & \sqrt{\frac{2}{2}} & 0 & \sqrt{\frac{3}{2}} & \ddots & \vdots \\ \vdots & \ddots & \sqrt{\frac{3}{2}} & 0 & \ddots & 0 \\ \vdots & \ddots & \ddots & \ddots & 0 & \sqrt{\frac{\mathrm{Dim}}{2}}  \\ 0      & \cdots & \cdots & 0      & \sqrt{\frac{\mathrm{Dim}}{2}}  & 0\end{bmatrix}\]
is the well known Jacobi matrix constructed with Hermite function recurrence relations \eqref{eqn:Recur}. 
\bibliographystyle{elsarticle-num} 

\end{document}